\documentclass[12pt]{iopart}

\usepackage{amsmath,amssymb,amsfonts}
\usepackage{algorithmic}
\usepackage{graphicx}
\usepackage{textcomp}

\usepackage[utf8]{inputenc}
\usepackage{mathrsfs}
\usepackage{orcidlink}
\usepackage{multirow}
\usepackage{subcaption}
\usepackage{cite}

\usepackage{xcolor}

\newcommand{\response}[1]{{#1}}
\newcommand{\resps}[1]{{#1}}
\newcommand{\respm}[1]{{#1}}

\begin{document}
\title[EIT meets ROM: a new framework]{Electrical Impedance Tomography meets Reduced Order Modelling: a framework for faster and more reliable electrical conductivity estimations}

\author{Matthew R. Walker \orcidlink{0000-0002-1175-2430}, Mariano Fern\'andez-Corazza \orcidlink{0000-0002-9624-5023}, Sergei Turovets \orcidlink{0000-0001-6956-0389} and Leandro Beltrachini \orcidlink{0000-0003-4602-1416}.}

\vspace{10pt}
\address{Matthew R. Walker and Leandro Beltrachini are with Cardiff University Brain Research Imaging Centre (CUBRIC), School of Physics and Astronomy, Cardiff University, UK.}
\address{Mariano Fern\'andez-Corazza is with the LEICI Institute of Research in Electronics, Control and Signal Processing, National University of La Plata, CONICET, Argentina.}
\address{Sergei Turovets is with NeuroInformatics Center, University of Oregon, Eugene, Oregon, USA.}



\begin{abstract}
\textbf{Objective:}~Inclusion of individualised electrical conductivities of head tissues is crucial for the accuracy of electrical source imaging techniques based on electro/magnetoencephalography and the efficacy of transcranial electrical stimulation.
Parametric electrical impedance tomography (pEIT) is a method to cheaply and non-invasively estimate them using electrode arrays on the scalp to apply currents and measure the resulting potential distribution. 
Conductivities are then estimated by iteratively fitting a forward model to the measurements, incurring a prohibitive computational cost that is generally lowered at the expense of accuracy. Reducing the computational cost associated with the forward solutions would improve the accessibility of this method and unlock new capabilities.
\textbf{Approach:}~We introduce reduced order modelling (ROM) to massively speed up the calculations of these solutions for arbitrary conductivity values.
\textbf{Main results:}~We demonstrate this new ROM-pEIT framework using a realistic head model with six tissue compartments, with minimal errors in both the approximated numerical solutions and conductivity estimations. We show that the computational complexity required to reach a multi-parameter estimation with a negligible relative error is reduced by more than an order of magnitude when using this framework. 
\response{Furthermore, we illustrate the benefits of this new framework in a number of practical cases, including its application to real pEIT data from three subjects. \textbf{Significance:}~Results suggest that this framework can transform the use of pEIT for seeking personalised head conductivities, making it a valuable tool for researchers and clinicians.}

\end{abstract}


\break 
\section{Introduction}
\label{sec:introduction}
Characterising the electromagnetic activity in the brain is essential for understanding its function in health and disease. The preferred methods to measure this activity are electroencephalography (EEG) and magnetoencephalography (MEG), forming the foundation of electrical source imaging (ESI) techniques. 
ESI methods rely on computational models of head tissues including anatomical structure and physical properties such as the electrical conductivity map~\cite{michel_eeg_2019}. The use of realistic and individualised head models has been shown to greatly improve the accuracy of these methods~\cite{mccann_impact_2022,vatta_realistic_2010}. In addition, realistic models aid the optimal placing of electrodes and dosage planning in Transcranial Electrical Stimulation (TES)~\cite{hunold_review_2023}. To generate these models, anatomical structure can generally be obtained from magnetic resonance or computerised tomography images using existing tools~\cite{puonti_accurate_2020}. However, electrical conductivities are typically selected as a population average for each tissue. A recent analysis has shown that conductivity values in all human head tissues likely vary significantly between individuals, challenging these assumptions~\cite{mccann_variation_2019}. Moreover, studies have shown that inaccurate conductivity values lead to errors in source localisation in EEG/MEG~\cite{haueisen_influence_2002,pohlmeier_influence_1997,vorwerk_influence_2019,wolters_influence_2006,mccann_impact_2022} and current localisation in transcranial electrical stimulation (TES)~\cite{wagner_investigation_2014,mccann_does_2021}. Therefore, there exists a need to estimate these conductivities on an individual basis.

Parametric electrical impedance tomography (pEIT) is a relatively affordable and non-invasive method for estimating the conductivities of tissues in a human head~\cite{holder_electrical_2004}. Using an array of electrodes placed on the scalp, a small current is injected and extracted from a subset and the electrical potential is measured on the complementary set.
This technique seeks to estimate the conductivities of head tissues by simulating forward solutions for sets of parameters and tuning the set to best match the electrical potential measurements taken. This allows one to characterise an individualised conductivity field.
As this model becomes more detailed, the computational expense of these simulations increases, presenting a serious limitation for highly realistic models where pEIT can take days to complete on a standard PC.

The current best effort to address computational load is to reduce the number of solutions required for pEIT to converge by utilising a gradient assisted optimisation method~\cite{simini_effects_2016}. This approach has proven successful for estimating scalp and skull conductivities from $in \ vivo$ and synthetic measurements to a good level of accuracy~\cite{fernandez-corazza_skull_2018,fernandez-corazza_analysis_2013}. However, this method requires the additional calculation of a gradient in each iteration, which itself is computationally costly. Furthermore, estimating the conductivity of some tissues proves challenging. For example, the conductivity of the spongiform bone inside the skull has been estimated with a coefficient of variation as large as one~\cite{fernandez-corazza_skull_2018}. 

In this work, we apply reduced order modelling (ROM) directly to pEIT to alleviate the computational demand while simultaneously \response{allowing the use of alternative optimisation methods for a faster pEIT framework.} 
ROM is a method utilised to find approximate numerical solutions to a parameterised boundary value problem quickly and accurately~\cite{quarteroni_reduced_2016}. This process consists of a computationally intensive offline training phase and a real-time online phase. During the offline phase, a reduced order model is constructed using solutions to the boundary value problem at different points in a multi-dimensional parameter space. The online phase then utilises this model for real-time approximations of solutions for any set of parameters.
We show that this framework yields significant improvements in the speed of the estimation of all tissues in the head, assimilating the new capability to confidently estimate conductivities previously unreachable \response{with traditional approaches.}




\section{Methods}
\label{methods}
\subsection{Parametric EIT Formulation}
\label{eit_maths}

\begin{figure}[t!]
    \centering
    \includegraphics[width=0.7\linewidth,trim=150pt 130pt 290pt 50pt, clip]{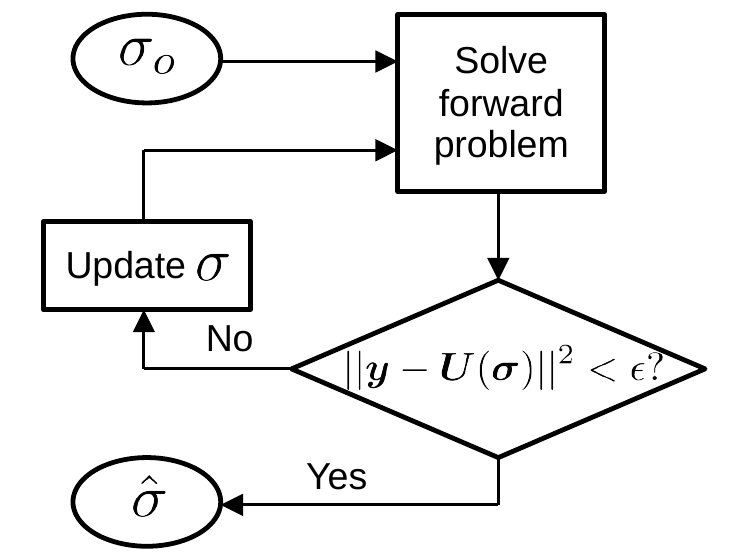}
    \caption{Flow chart of the traditional implementation of the inverse problem for pEIT. Here, $\epsilon$ refers to a stopping threshold, $\boldsymbol{\sigma_{0}}$ is the initial conductivity guess and $\boldsymbol{\hat{\sigma}}$ is the estimated conductivity value. Note that each loop requires at least a full calculation of the forward problem.}
    \label{fig:flowchart}
\end{figure}

Parametric EIT is an ill-posed inverse problem (IP) that results in estimates of the electrical conductivities of tissue compartments. 
This is done by iteratively minimising the squared error between the measurements $\boldsymbol y \in \mathbb{R}^{L}$ and the conductivity-dependent simulated signals $\boldsymbol U \in \mathbb{R}^{L}$ on $L$ electrodes. Mathematically, this is generally expressed as
\begin{equation}
    \boldsymbol{\hat{\sigma}} = \arg \min_{\boldsymbol{\sigma}}\{(\boldsymbol y - \boldsymbol U(\boldsymbol{\sigma}))^T(\boldsymbol y - \boldsymbol U(\boldsymbol{\sigma}))\} \text{,}
    \label{MLE}
\end{equation}
where 
$\boldsymbol{\hat{\sigma}}$ are the estimated conductivities~\cite{fernandez-corazza_analysis_2013}. 
This results in an optimisation process that requires the calculation of one or more forward problems (FPs) at each iteration and then updating $\boldsymbol \sigma$ based on the error and the optimisation technique used (Fig.~\ref{fig:flowchart}).

The pEIT-FP is a boundary value problem governed by a Laplace equation subject to Neumann boundary conditions~\cite{somersalo_existence_1992}. The formulation and numerical methods for the pEIT-FP solution are well documented in the literature~\cite{somersalo_existence_1992,fernandez-corazza_analysis_2013,fernandez-corazza_estimation_2011,fernandez-corazza_skull_2018,vauhkonen_three-dimensional_1999}.
Here, we adopt the finite element (FE) method due to its flexibility to handle arbitrarily-shaped compartments.
After discritisation, the variational formulation of the pEIT-FP considering the complete electrode model (CEM)~\cite{kuo-sheng_cheng_electrode_1989,somersalo_existence_1992} results in the system~\cite{vauhkonen_three-dimensional_1999}
\begin{equation}
    \boldsymbol A(\boldsymbol{\sigma}) \boldsymbol u(\boldsymbol{\sigma}) = \boldsymbol b \text{,}
    \label{CEM_sys}
\end{equation}
where,
\begin{subequations}
\label{CEM_sys_parts}
\begin{gather}
    \boldsymbol A(\boldsymbol{\sigma}) = \begin{bmatrix} \boldsymbol K(\boldsymbol{\sigma}) & - \boldsymbol B \\ - \boldsymbol B^T & \boldsymbol C\end{bmatrix} \text{,} \\ 
    \boldsymbol u(\boldsymbol{\sigma}) = \begin{bmatrix}\boldsymbol u_n(\boldsymbol{\sigma}) \\ \boldsymbol U(\boldsymbol{\sigma})\end{bmatrix} \text{,} \ 
    \boldsymbol b = \begin{bmatrix}\boldsymbol 0 \\ \boldsymbol I \end{bmatrix} \text{,}
\end{gather}
\end{subequations}
$u_n(\boldsymbol{\sigma}) \in \mathbb{R}^n$ is the solution vector on the $n$ nodes of the volumetric FE mesh and $\boldsymbol I \in \mathbb{R}^L$ is the vector of injection currents on the electrodes. The matrix $\boldsymbol{K}(\boldsymbol{\sigma}) \in \mathbb{R}^{n \times n}$ is known as the stiffness matrix and depends on the conductivity values of each compartment $\boldsymbol{\sigma} = \{\sigma_1,\sigma_2,...,\sigma_P\}$, where $P$ is the number of tissue compartments. The matrices $\boldsymbol B \in \mathbb{R}^{n \times L}$ and $\boldsymbol C \in \mathbb{R}^{L \times L}$ encode information about the electrodes on the surface of the domain and do not depend on the conductivity. The entries of the matrices $\boldsymbol K$, $\boldsymbol B$ and the diagonal matrix $\boldsymbol C$ are given by~\cite{vauhkonen_fixed-lag_2001}
\begin{subequations}
\label{a_stiffness}
\begin{gather}
K_{ij} = \int_{\Omega} \langle \sigma\nabla\psi_i,\nabla\psi_j \rangle d\Omega + \sum^L_{l=1}\frac{1}{z_l} \int_{e_l} \psi_i \psi_j d(\partial\Omega) \text{,} \label{a_stiffa}\\
B_{il} = \frac{1}{z_l}\int_{e_l}\psi_id(\partial\Omega) \text{,} \label{a_stiffb}\\
C_{ll} = \frac{1}{z_l}\int_{e_l}d(\partial\Omega) = \frac{|e_l|}{z_l} \textrm{,} \label{a_stiffc}
\end{gather}
\end{subequations}
where $e_l$ represents the $l$th electrode, $|e_l|$ its area, $z_l$ its contact impedance, 
$\Omega$ is the domain (i.e., the head) with boundary $\partial\Omega$, and $\psi_i$ is a basis function on the nodes $i = 1,2,...,n$.


A useful property of the matrix $\boldsymbol K(\boldsymbol{\sigma})$ is that, in the case of homogeneous conductivities, it can be linearly decomposed into several constituent stiffness matrices $\boldsymbol{K}_p \in \mathbb{R}^{n \times n}$, each representing a different compartment $p$ in the head model and independent of $\boldsymbol \sigma$~\cite{fernandez-corazza_analysis_2013}. Consequently, the matrix $\boldsymbol{A}(\boldsymbol{\sigma})$ can be split into $P+1$ matrices $\boldsymbol{A}_p \in \mathbb{R}^{(n+L) \times (n+L)}$, i.e.,
\begin{equation}
    \boldsymbol{A}(\boldsymbol{\sigma}) = \boldsymbol{A}_0 + \sum_{p=1}^{P} \sigma_p \boldsymbol{A}_p \text{,}
    \label{stiff_assem}
\end{equation}
where $\boldsymbol{A}_0$ is a $\boldsymbol{\sigma}$-independent matrix encoding the information from matrices $\boldsymbol B$ and $\boldsymbol C$ and the second term in eq.~\eqref{a_stiffa}. 
It is straightforward to show that such a decomposition holds 
even in the case of anisotropic conductivities~\cite{fernandez-corazza_analysis_2013}.
This property is referred to as \textit{affine decomposition} of the parameters of interest (i.e., the conductivities) and it is a fundamental requirement for a system where ROM is applied. 

\subsection{Reduced Order Modelling}
\label{rom_maths}

ROM is a mathematically rigorous technique to efficiently build a low-dimensional model mapping changes in a set of conductivities to changes in the solution of eq.~\eqref{CEM_sys}~\cite{quarteroni_reduced_2016}. This model is constructed in an offline phase using a relatively small number of $N<<(n+L)$ strategically selected solutions of eq.~\eqref{CEM_sys} with specific conductivities, which are then used in the `online' phase to find rapid solutions for any set of conductivities. Below, we present a brief overview of the fundamental principles of ROM.

Taking advantage of the affine decomposition, massively reduced versions of the $\boldsymbol A_p$ matrices can be formed using the reduced model, allowing the assembly of a reduced system in the online phase at any point in the $P$-dimensional parameter space $\mathscr{P} \in \mathbb{R}^P$ (i.e., for any set of conductivities). This new system can be solved in real-time, resulting in a reduced-basis solution $\boldsymbol u_{N} \in \mathbb{R}^N$ that is easily transformed to $\boldsymbol u_{a} \in \mathbb{R}^{n+L}$ approximating the solution of the high-dimensional system $\boldsymbol u$.

The model is trained using a number of full-order solutions, called snapshots, which are selected strategically across $\mathscr{P}$. Judiciously choosing the points with which to build the reduced model is done by employing a greedy algorithm. 
A distinguishing feature of ROM is the presence of a rigorous upper bound $\Delta(\boldsymbol \sigma)$ on the error of the approximate solutions, which guides the greedy algorithm in the snapshot selection, acting as a proxy for the error~\cite{quarteroni_reduced_2016}.
This bound on the error can be calculated almost instantly for any given point in $\mathscr{P}$ and can therefore efficiently explore the space to guide the next snapshot point.
During each iteration of the greedy algorithm, the bound is calculated for a finite sample set $\Xi \subset \mathscr{P}$ and a snapshot is generated using the conductivity set that minimises it. $\Xi$ is chosen to represent the entire P-dimensional space~$\mathscr{P}$.
Utilising the bound to select the snapshots presents two advantages. Firstly, it allows an extremely quick assessment of the maximum error attainable at a fine discritisation of $\mathscr{P}$.
Secondly, it can be used as a stopping criterion for certifying the maximum error in $\boldsymbol u_{a}$~\cite{quarteroni_reduced_2016}.
The relationship between the $a \ posteriori$ relative error [RE($\boldsymbol \sigma$)] for a given point in $\mathscr{P}$ and the $a \ posteriori$ relative error bound [$\Delta_{RE}(\boldsymbol \sigma)$] is~\cite{quarteroni_reduced_2016}
\begin{equation}
\text{RE}(\boldsymbol \sigma) \triangleq \frac{||\boldsymbol u(\boldsymbol \sigma)-\boldsymbol u_{a}(\boldsymbol \sigma)||_{L_2}}{||\boldsymbol u_{a}(\boldsymbol \sigma)||_{L_2}} \leq \frac{\Delta(\boldsymbol \sigma)}{||\boldsymbol{u}_{N}(\boldsymbol \sigma)||_{L_2}} \triangleq \Delta_{RE}(\boldsymbol \sigma)\text{.}
\label{bound1}
\end{equation}

\begin{figure}[t!]
    \centering
    \includegraphics[width=0.7\linewidth,trim=40pt 200pt 70pt 100pt, clip]{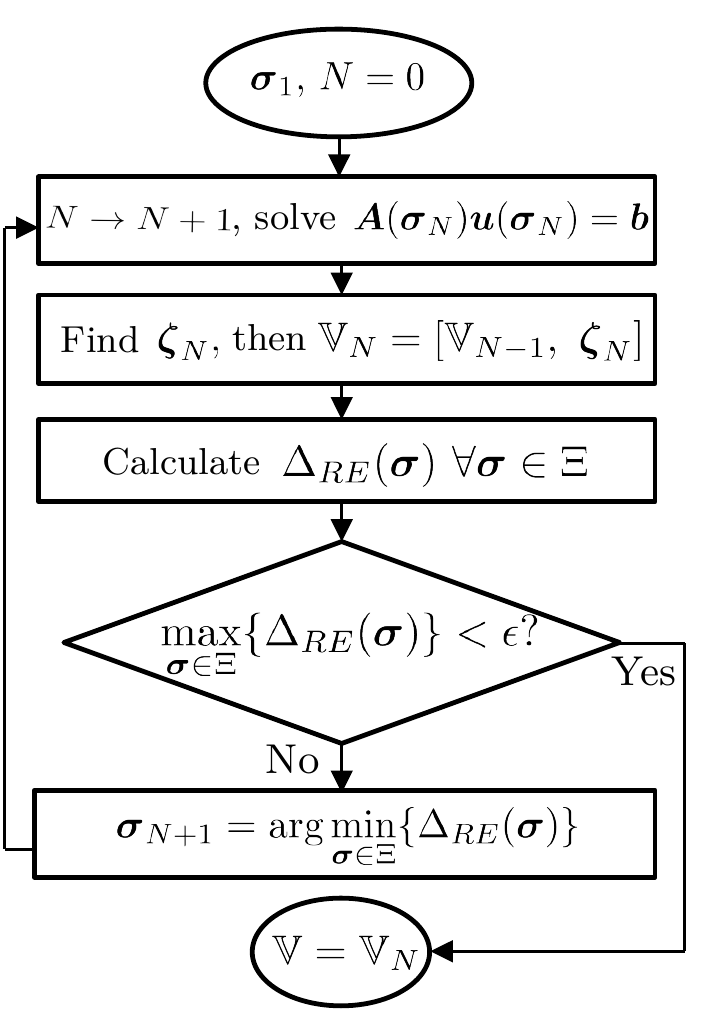}
    \caption{Greedy algorithm used in the offline training phase for ROM where $\epsilon$ is some stopping threshold (see text).}
    \label{fig:flowchart_ROM}
\end{figure}

The reduced model takes the form of a reduced-basis space, built using the snapshots calculated by the greedy algorithm. To obtain the reduced system, the full-order stiffness matrices are projected on the space during the offline phase.
This reduced-basis space is represented by the matrix ${\mathbb{V} \in \mathbb{R}^{(n+L) \times N}}$.
To construct the orthonormal basis $\mathbb{V}$ we perform a Gram-Schmidt orthonormalisation on a snapshot, before adding it to the orthonormal basis iteratively. We begin by selecting a random parameter vector $\boldsymbol \sigma_1 \in \Xi$ and computing the full-order solution $\boldsymbol{u}(\boldsymbol \sigma_1)$. 
The first basis vector for the orthonormal space is simply the first snapshot, which is a full-order solution (i.e., $\boldsymbol \zeta_1(\boldsymbol \sigma_1) = \boldsymbol{u}(\boldsymbol \sigma_1)$). Thereafter, the orthonormalised solutions $\boldsymbol \zeta_j(\boldsymbol \sigma)$ for the $j$th snapshot are concatenated,
\begin{equation}
\mathbb V = [\boldsymbol{u}(\boldsymbol \sigma_1),\boldsymbol \zeta_2(\boldsymbol \sigma_2),...,\boldsymbol \zeta_{N}(\boldsymbol \sigma_N)] \text{,}
\end{equation}
such that $\{\boldsymbol \sigma_1, \boldsymbol \sigma_2,...,\boldsymbol \sigma_N\} \subset \Xi$.
Also known as the transformation matrix, $\mathbb V$ relates the projected stiffness matrix $\boldsymbol{A}_{N}(\boldsymbol \sigma) \in \mathbb{R}^{N \times N}$ and projected independent vector $\boldsymbol{b}_{N}(\boldsymbol \sigma) \in \mathbb{R}^{N}$ with the full-order versions through the expressions~\cite{quarteroni_reduced_2016}
\begin{equation}
\boldsymbol{A}_{N}(\boldsymbol \sigma) = \mathbb{V}^T \boldsymbol{A}(\boldsymbol \sigma)\mathbb{V} \text{,} \ \ \boldsymbol{b}_{N}(\boldsymbol \sigma) = \mathbb{V}^T \boldsymbol{b}(\boldsymbol \sigma) \textrm{,}
\label{translation}
\end{equation}
resulting in the reduced system to solve
\begin{equation}
\boldsymbol A_{N}(\boldsymbol \sigma) \boldsymbol u_{N}(\boldsymbol \sigma) = \textbf{b}_{N}(\boldsymbol \sigma)\textrm{,}
\label{system_rb}
\end{equation}
where $\boldsymbol u_{a}(\boldsymbol \sigma) = \mathbb{V} \boldsymbol u_{N}(\boldsymbol \sigma)$.
It is clear from eq.~\eqref{translation} that, as $N<<(n+L)$, the dimensions of the 
resulting system are massively reduced, requiring significantly fewer operations to solve.
Ultimately, this means that a FP can be calculated at any point in $\mathscr{P}$ almost instantly.
Fig.~\ref{fig:flowchart_ROM} shows a flowchart of the greedy algorithm, demonstrating the construction of $\mathbb{V}$.

Finally, it should be noted that the calculation of the bound relies on a $\boldsymbol{\sigma}$-dependent parameter called the stability factor $\beta_h(\boldsymbol \sigma)$, related in the following way
\begin{equation}
    \Delta(\boldsymbol \sigma) = \frac{||\boldsymbol{b} - \boldsymbol{A(\sigma)}\mathbb{V}\boldsymbol{u}_N(\boldsymbol{\sigma})||_{L_2}}{\beta_h(\boldsymbol \sigma)} \ \textrm{.}
    \label{bound2}
\end{equation}
The numerator of eq.~\eqref{bound2} is known as the residual and can be found very quickly with some computational splitting inside the greedy algorithm.
Obtaining the stability factor, however, is a more computationally intensive calculation requiring the solution to a generalized eigenvalue problem~\cite{quarteroni_reduced_2016}. Therefore, we employ a similar schema as before, splitting it into an offline training phase and online real-time phase. The offline phase involves creating an interpolant using radial basis functions and interpolation points in $\mathscr{P}$ which can then be used in the online phase for a quick evaluation of $\beta_h(\boldsymbol \sigma)$ for any point in $\mathscr{P}$.
For details on the splitting of the residual, calculation of the bound, its offline/online decomposition, and its calculation for a rank-deficient stiffness matrix, the reader is referred to Quarteroni et al. (2016)~\cite[Chs.3,4,6]{quarteroni_reduced_2016}.

\subsection{Implementation and Experiments}

\subsubsection{Set-up}
\label{setup}
\begin{figure}[t!]
    \centering
    \hspace*{0cm}
    \includegraphics[width=1\linewidth,trim=0pt 100pt 0pt 150pt, clip]{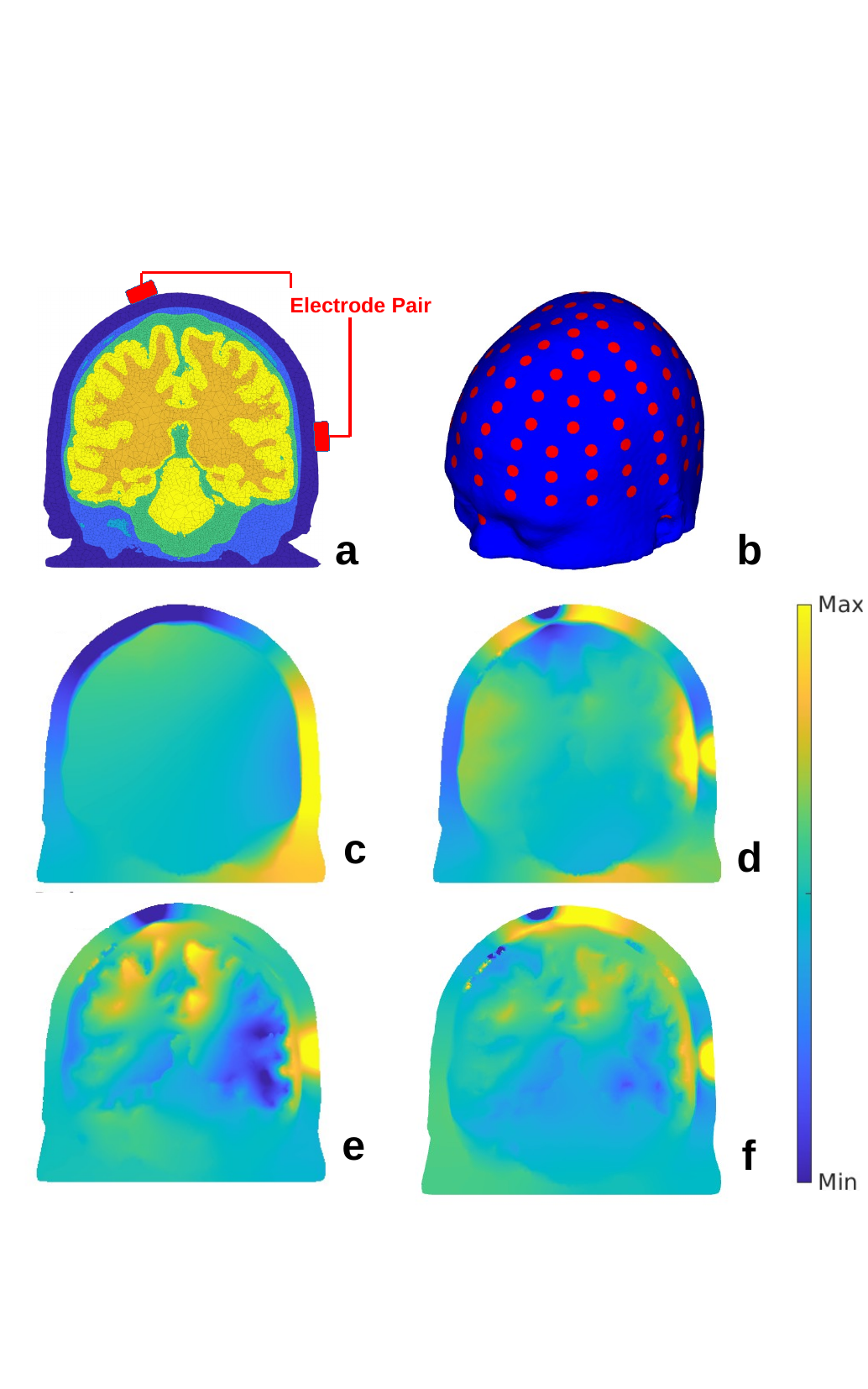}
    \caption{\response{\textbf{a}) Cross section of the FE mesh with compartments coloured separately. \textbf{b)} Electrodes (red circles) modelled on the surface of the scalp. \textbf{c-f)} Selection of the first basis vectors ($\boldsymbol{\zeta}_2$, $\boldsymbol{\zeta}_4$, $\boldsymbol{\zeta}_{7}$ and $\boldsymbol{\zeta}_{11}$, respectively) for the transformation matrix $\mathbb{V}$  made for an electrode pair plotted on the FE mesh. The colour indicates the value of the projected basis vector at each node that represents the additional information being encoded.}}
    \label{fig:training}
\end{figure}

We used a realistic head model discritised with 5M tetrahedral elements and 800k nodes. The model was based on the Colin27 atlas~\cite{aubert-broche_new_2006} and processed as in previous publications~\cite{mccann_impact_2022}. A cross section is shown in Fig.~\ref{fig:training}a depicting different tissue compartments, i.e., scalp, compact skull bone, spongiform bone, cerebrospinal fluid (CSF), grey matter (GM) and white matter (WM). 
The conductivities chosen for the synthetic measurements were uniform random samples within the \response{interquartile ranges} described in Table~\ref{table:conds} for each of the tissues. The \response{minimum, lower and upper quartiles and maximum values (excluding outliers) were chosen from} the work carried out by McCann et al. (2019)~\cite{mccann_variation_2019}. A reduced model for each electrode pair used was trained for conductivity parameters within these ranges.

\begin{table}[b]
\centering
\begin{tabular}{||c|c@{\hspace{1em}} c@{\hspace{1em}} c@{\hspace{1em}} c@{\hspace{1em}} c@{\hspace{1em}} c||}
 \cline{2-7}
  \multicolumn{1}{c|}{} & Scalp & Compact & Spongiform & CSF & GM & WM\\ 
  \multicolumn{1}{c|}{} & & bone & bone & & & \\ [0.5ex]
  \hline
 Min. (S/m) & 0.136 & 0.0008 & 0.001 & 1.388 & 0.060 & 0.065 \\ 
 LQ (S/m) & 0.303 & 0.002 & 0.013 & 1.450 & 0.268 & 0.092 \\ 
 UQ (S/m) & 0.444 & 0.009 & 0.043 & 1.794 & 0.508 & 0.177 \\
 Max. (S/m) & 0.620 & 0.0131 & 0.088 & 1.794 & 0.739 & 0.228 \\
 \hline
\end{tabular}
\caption{Ranges of conductivities used for training. Minimum, lower quartile (LQ), upper quartile (UQ),  and maximum (excluding outliers) reported by McCann et al. (2019).} 
\label{table:conds}
\end{table}

\subsubsection{Technical Implementation}

For each conductivity sample, the FP was solved for each of the 132 pairs of electrodes, where the injection and extraction electrode had $20 \ \mu A$ and $-20 \ \mu A$ current applied, respectively. All pairs are composed of a unique injection electrode and a sink electrode that is common for all pairs placed on the scalp above the Sagittal suture \resps{(position Cz). This choice of electrode pairs allows the flexibility to simulate any electrode pair through a simple linear combination of the solutions of the trained pairs.} \response{Electrodes were positioned according to the ABC-128 standard layout (as used in BioSemi products) with the addition of fiducial electrodes placed in the nasion, inion and left and right preauricular points and an electrode at the vertex (Cz), resulting in 133 sensors. This layout is partially displayed in Fig.~\ref{fig:training}b. An average common reference (AR) was applied to the potentials on the electrodes.}

The systems of equations were solved with the Preconditioned Conjugate Gradient (PCG) solver with incomplete LU preconditioners~\cite{elman_finite_2014}. They were solved with a tolerance of $10^{-10}$ and a maximum number of iterations of 6000. The Gaussian noise added to the measurements had a standard deviation of $0.82 \ \mu V$, which is similar to the noise found in real measurements~\cite{fernandez-corazza_skull_2018}. The 133 electrodes were modelled as $1\ cm$ diameter circles on the surface of the scalp with an effective contact impedance of $5 \ \Omega m^2$.

The FE method was implemented using first-order linear basis functions on the mesh nodes as used by Vauhkonen et al. (1999)~\cite{vauhkonen_three-dimensional_1999}. 
Analytical expressions of the element matrices needed in eqs.~\eqref{a_stiffa}-\eqref{a_stiffc} were utilised to avoid errors due to numerical quadrature~\cite{beltrachini_analytical_2019}.

The ROM method was trained using the same model, injection patterns and range of conductivities as above. We chose to train ROM for up to 100 snapshots to demonstrate the reduction in error in the FPs and IPs. However, as will become clear, there are a number of stopping criteria that can guide how many snapshots to take.

Similarly to other work~\cite{fernandez-corazza_skull_2018}, we have removed some erroneous estimations from injection patterns where the IP has either not converged or has given an unrealistic conductivity (e.g., negative conductivities), which may occur for the traditional method only as it is based on an unconstrained optimisation technique.

\response{It should be noted that the matrix $\boldsymbol{A}_0$ in eq.~\eqref{stiff_assem} can be further affinely decomposed into a impedance (``$z$") independent matrix with a coefficient equal to $z^{-1}_l, \ l=1,...,L$. In this way, $z_l$ can be trained as an additional set of parameters. However, we found that even across a large range of contact impedances, the effect on the FPs and IPs was negligible, as reported for EEG~\cite{pursiainen_complete_2012}.}

\response{The reduced models generated for each electrode pair are completely independent, similarly to the IPs for the traditional method. Therefore, computational work was trivially parallelised by electrode pair on a cluster computer with 11 Intel(R) Xeon(R) X5660 CPU nodes at 2.80GHz. Each node had 12 cores and 16GB of memory per core.}

\subsubsection{Experiment 1 - ROM Performance}

Our first experiment serves two main purposes. The first is to confirm that the pEIT-FP is meaningfully reducible in the sense that, for small $N$ values, $\boldsymbol{u}_{a}$ quickly converges to $\boldsymbol u$. The second is to validate our bound while simultaneously assessing its tightness.
To achieve these aims, we plotted the average and maximum RE($\boldsymbol \sigma$) and $\Delta_{RE}(\boldsymbol \sigma)$ as a function of $N$.
The $\Delta_{RE}(\boldsymbol \sigma)$  was calculated in the training phase during the greedy algorithm for a 6000 sample train across $\mathscr{P}$ for each electrode pair. The mean and maximum $\Delta_{RE}(\boldsymbol \sigma)$  across the sample train were found for each electrode pair and then averaged across all of them. 
The RE($\boldsymbol \sigma$) was calculated for each electrode pair for 100 samples of $\mathscr{P}$. The average RE($\boldsymbol \sigma$) across all electrodes for each sample was found before plotting the average and maximum across $\mathscr{P}$. This was repeated for an increasing number of snapshots.

\subsubsection{Experiment 2 - IP Performance}
To assess how useful the ROM-pEIT framework is, we considered two important metrics: the accuracy of the IP solutions and the computational cost required to achieve them. To that end, we compared our results with the best approach currently in the field, which provides reliable estimations for scalp and compact skull electrical conductivities~\cite{fernandez-corazza_skull_2018}. This method minimises eq.~\eqref{MLE} using the gradient-assisted quasi-Newton method. However, this requires the calculation of the gradient of the solution for each FP, for each of the parameters being searched for~\cite{dennis_numerical_1996, fernandez-corazza_skull_2018}. The gradient can be found using~\cite{fernandez-corazza_skull_2018}
\begin{equation}
    \frac{\partial \textbf{A}^{-1}(\boldsymbol \sigma)\textbf{b}}{\partial \sigma_p} = -\textbf{A}^{-1}(\boldsymbol \sigma)\textbf{A}_p\boldsymbol{u} \text{,} \ \forall p=1,...,P.
    \label{derivative}
\end{equation}
From eq.~\eqref{derivative}, it is clear that finding each of the gradients requires solving another large system of equations similar to the FP. This results in a significant overhead in terms of computational cost, especially when multiple parameters are being estimated simultaneously. Inserting this into the loop in Fig.~\ref{fig:flowchart} shows that, for each iteration in the optimisation, the number of large systems of equations to solve is equal to the FP plus the number of tissues being estimated.
Henceforth, we shall refer to this method of gradient assisted optimisation using the full-order FP as the traditional method.

A further consequence of using the reduced system of eqs.~\eqref{system_rb} is that the derivative~\eqref{derivative} can no longer be calculated and therefore neither can the quasi-Newton method be utilised efficiently.
However, using quasi-Newton methods to reduce the computational cost is no longer of concern, and we are free to explore other methods, such as the interior-point optimisation approach. Although this method requires more loops and therefore more systems to solve than the quasi-Newton algorithm, the cost of the new optimisation is still negligible compared to the traditional technique. Therefore, we have chosen to compare the computational cost of the ROM-pEIT framework and the traditional method by using the number of $(n+L) \times (n+L)$ linear systems of equations needed to be solved for each electrode pair. For ROM, all of these systems are solved in the offline phase. Given that these systems embody the bulk of the computational work, it is an appropriate metric for comparison. Making the comparison independent of the electrode pairs means that the savings are the same irrespective of the injection protocols used.

For the traditional method, the IP was run as a 3-parameter search, optimising for the scalp, compact skull and spongiform bone simultaneously. For the conductivities not being optimised (CSF, GM, WM), they were fixed to the reference truth values used to make the synthetic measurements. We chose this format to isolate and assess the estimation of the three conductivities stated only.

\response{To assess the improvement in the IP, we redefined the relative error (RE) as }

\response{\begin{equation}
    RE = |\hat{\sigma} - \sigma|/\sigma \text{,}
\end{equation}}

\response{\noindent where $\hat{\sigma}$ and $\sigma$ are the estimated and the ground truth scalar conductivities, respectively.}
The estimation progress was logged at each iteration and plotted as the RE between the estimation and the sample parameters. The ROM IP was run as a 6-parameter search to estimate all of the compartments in the model. 
All optimisations were started from the centre point of the ranges specified in Table~\ref{table:conds}.

The mean of the RE in the estimation for each tissue for each number of iterations (and function evaluations within those iterations) was calculated, and then averaged across 10 randomly selected conductivity samples. We used 10 samples due to the computational cost of the traditional method. The IP with ROM was then run for a further 90 samples of $\mathscr{P}$ and plotted separately with the average RE across the samples and electrodes displayed for all tissues.

\response{To further assess the estimations, we repeated the ROM-pEIT IP with the reduced model containing 30 snapshots for various signal-to-noise ratios by increasing the standard deviation in the additive Gaussian noise. The RE in each tissue was averaged across all electrode pairs and samples.}



\subsubsection{Experiment 3 - Anisotropy}
\label{method:anis}

It has been shown that the inclusion of the spongiform bone in head models reduces the error in the EEG-FP and IP~\cite{mccann_impact_2022}. 
However, in the event of missing spongiform information, the skull may be modelled as a single compartment with anisotropic conductivity~\cite{fernandez-corazza_analysis_2013,dannhauer_modeling_2011}. Therefore, a separate experiment aimed to demonstrate the adaptation of ROM-pEIT to model a homogeneous and anisotropic skull conductivity.


Firstly, we modified the realistic head model by merging the compact and spongiform bone to create one homogeneous skull compartment. We then trained another ROM model with the new head model where the conductivity tensor field for the skull compartment has been transformed from a Cartesian basis to a radial and tangential basis relative to the centre point of the brain. The range of values used for both radial and tangential conductivities were from the minimum compact skull (0.002 S/m) to the maximum spongiform skull (0.043 S/m) used in the previous experiments. This was to accommodate for a wide range of possible skull compositions, from entirely compact skull to significant proportions of spongiform bone.

We analysed the sensitivity of the ROM-pEIT framework to anisotropic conductivities in the skull by assessing the RE in each compartment. To achieve this, we created 100 synthetic measurements using the full-order model with noise. The model was adapted by merging the compact and spongiform skull and given an anisotropic conductivity in the same range used for ROM. These measurements were then utilised to run the IP with a new reduced model, trained with radial and tangential conductivities in the whole skull. We plotted the RE in the estimation for each tissue compartment to assess the sensitivity of the reduced basis IP to the radial and tangential components of the skull conductivity. As before, the IP was run as a 6-parameter estimation, this time estimating the radial and tangential values, replacing the compact and spongiform skull conductivities.


\subsubsection{Experiment 4 - Response to Reference Choice}

\response{A common reference electrode is standard in EEG setups to ensure unbiased readings between amplifiers. This electrode is most often fixed in place at the vertex (electrode Cz). However, it has been suggested that a more flexible reference could yield fewer artifacts in EEG data~\cite{hari_meg-eeg_2017}. Although choosing a reference for the potential measurements has been investigated thoroughly through the lens of ``re-referencing" techniques in EEG, to our knowledge, no such analysis has been performed for pEIT. Furthermore, although re-referencing allows the use of average reference (AR) and the reference electrode standardisation technique (REST), these methods still encode the information of the originally referenced potentials and, as such, are impacted heavily by the use of realistic head models~\cite{chella_impact_2016,liu_estimating_2015,yao_method_2001}. In this context we used ROM-pEIT to investigate the impact of the original choice of reference electrode in a situation of missing spongiform bone information.}

\response{Firstly, we trained a reduced model with the compact and spongiform bone merged into a single whole skull compartment. The conductivity ranges used in training for each tissue were the interquartile ranges in Table~\ref{table:conds} where the whole skull compartment was trained between the lower quartile for compact bone and upper quartile for spongiform bone. 
\resps{We created a current injection protocol comprising 106 injection-extraction electrode pairs with a single common reference. These pairs included a set of 106 unique injection electrodes, where the extraction electrodes were selected to be the opposite side of the head (with electrodes repeating no more than twice). This strategy was chosen to obtain maximum coverage of the head and probe deep tissues~\cite{fernandez-corazza_analysis_2013}. We repeated the protocol sequentially changing the reference to every electrode position (and omitting pairs involving such an electrode). This resulted in 133 sets of synthetic measurements with a unique reference electrode in each set. Next, we ran the IP for each electrode pair in the protocol in all sets, totalling over 14,000 IPs.}}

\resps{To assess the effect of the reference selection on the estimations, we monitored the standard deviation in the IP skull estimations in each set and then plotted this value on the reference electrode of that set, before interpolating across the skull. We then reported the estimations and standard deviations for the reference electrode that has the most variability.}



\subsubsection{Experiment 5 - Validation with Real Data}

\response{We use real pEIT data from 44, 46 and 52 year old male subjects labelled AM (Atlas Man), CA (Caucasian Atlas) and AA (Asian Atlas), respectively. The head tissues for these subjects were segmented from a T1-weighted MRI co-registered with a CT scan and the FEM models were generated with the \textit{iso2mesh} package~\cite{qianqian_fang_tetrahedral_2009}. The data was acquired using a 128 sensor net from EGI (Electrical Geodesics, Inc.) with one reference electrode (Cz) where 62 unique injection patterns were applied using a current of $\pm$20~$\mu A$ at a frequency of 27 Hz. Further details on the image processing and data acquisition are described in Fern\'andez-Corazza et al. (2018)~\cite{fernandez-corazza_skull_2018}. The data was cleaned by removing measurements from 3 bad channels and removing patterns whose data was also noisy. This resulted in 36, 47 and 42 patterns with usable data, respectively, each with one injection, one extraction and 123 measurement electrodes. The injection and extraction electrodes were approximately diametrically opposite sides of the head. \par All research protocols involving human subjects complied with the ethical standards in the Helsinki Declaration of 1975 and approved by EGI’s Institutional Review Boards (IRB). Informed consent was obtained for each subject. \par}
\response{The ranges used to train the reduced models for each subject were initially expanded to the minimum and maximum conductivities given in Table~\ref{table:conds}. We then trained secondary reduced models with the conductivity range for spongiform bone expanded to [0.001, 0.3] S/m to include the values estimated by Fern\'andez-Corazza et al. (2018).}

\response{To replicate those results, we first ran the IP considering the scalp, compact bone and spongiform bone compartments to be estimated and all others fixed to the same values used in~\cite{fernandez-corazza_skull_2018} for both ranges of conductivities. We then ran the IP considering all compartment conductivities to be estimated simultaneously, for the ranges given in Table~\ref{table:conds}.} 



\begin{figure}[t]
    \centering
    \hspace*{-0cm}
    \includegraphics[width=1\linewidth,trim=50pt 1pt 40pt 20pt, clip]{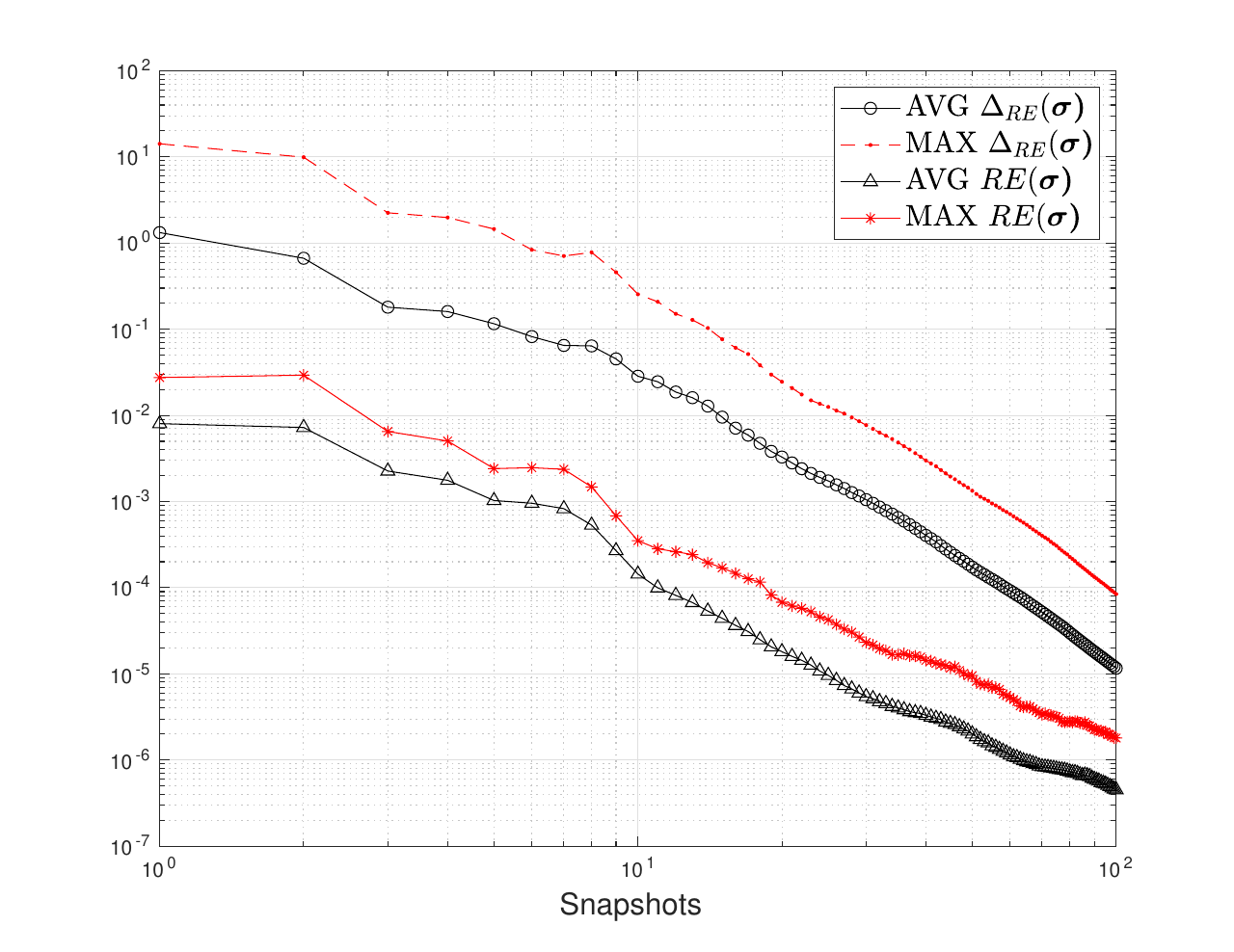}
    \caption{Average and maximum $\Delta_{RE}(\boldsymbol \sigma)$ and RE$(\boldsymbol{\sigma})$ for a sample of parameters (averaged across electrodes) against the number of snapshots.} 
    \label{fig:bound}
\end{figure}


\section{Results}
\label{apps}

\begin{figure}[t]
    \centering
    \includegraphics[width=1\linewidth,trim=20pt 30pt 60pt 50pt, clip]{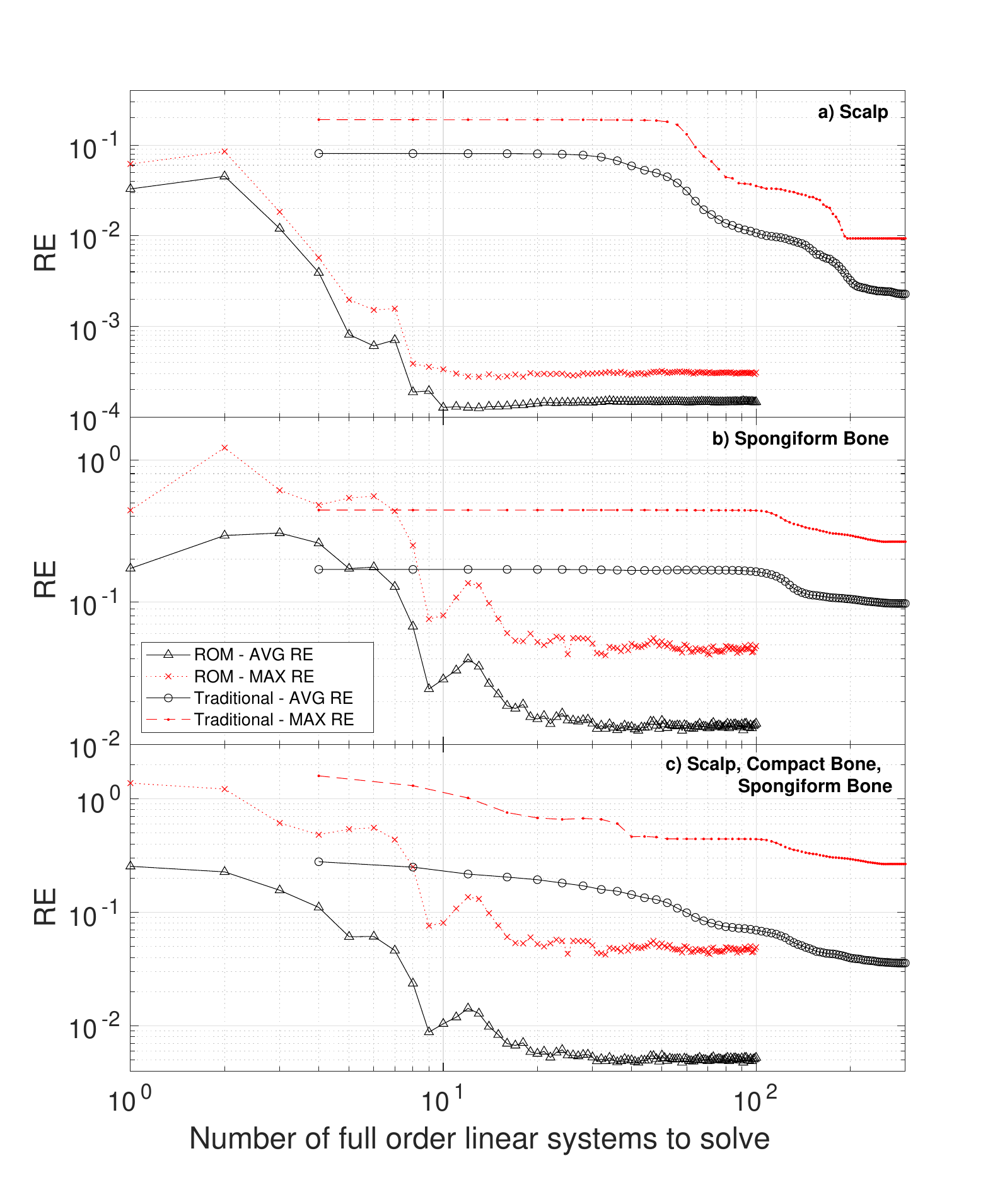}
    \caption{Average (black lines) and maximum (red lines) of the RE in the estimation of the conductivities across multiple electrode pairs and for 10 sets of synthetic measurements with uniformly distributed conductivity samples. The red and black dotted lines in each figure correspond to the traditional method and the red and black full lines with crosses and triangles respectively are for ROM.} 
    \label{fig:inv_snap_all}
\end{figure}


\subsubsection{Experiment 1 - ROM Performance}

Figs.~\ref{fig:training}c-f show a subset of the basis vectors (i.e., $\boldsymbol{\zeta}_i$ for $i=2,4,7,11$) that constitute the reduced basis space. Each additional function to the first is an orthogonal projection to the matrix $\mathbb{V}$ and encodes additional information into the reduced model. 
In particular, the basis vector $\boldsymbol{\zeta}_{11}$ (Fig.~\ref{fig:training}f) shows that after the projection there is a significant difference in electrical potential solution in the brain between the previous sample conductivities and those for the snapshot. Once added, this results in a reduced model with specific information about the response of the electrical potential in the brain to conductivity changes in the model. 
This demonstrates the greedy algorithm in action. The same effect can be seen with the spongiform bone with respect to the bright spots in the skull in basis vectors $\boldsymbol{\zeta}_{4}$ and $\boldsymbol{\zeta}_{11}$.

Fig.~\ref{fig:bound} shows the average and maximum $\Delta_{RE}(\boldsymbol \sigma)$ and RE$(\boldsymbol \sigma)$ as a function of snapshots. The $\Delta_{RE}(\boldsymbol \sigma)$ was calculated across the sample set $\Xi$ and the RE$(\boldsymbol \sigma)$ was found for 100 conductivity samples.
It is interesting to note that the bound becomes slightly sharper as the number of snapshots increases. 
Fig.~\ref{fig:bound} also demonstrates that $\Delta_{RE}(\boldsymbol \sigma)$ can be used as a stopping criteria for the number of snapshots used to train the model. After being set, the greedy algorithm will stop when $\Delta_{RE}(\boldsymbol \sigma)$ for every point in the fine sample is below the threshold stated. Using this stopping criteria ensures that the RE($\boldsymbol \sigma$) in the FP is below the threshold. However, choosing a threshold is not trivial (see Section~\ref{discussion}) and there is a risk of unnecessary training of the model. 

\subsubsection{Experiment 2 - IP Performance}
\label{inv_perf}

\begin{figure}[t]
    \includegraphics[width=\linewidth,trim=20pt 10pt 50pt 30pt, clip]{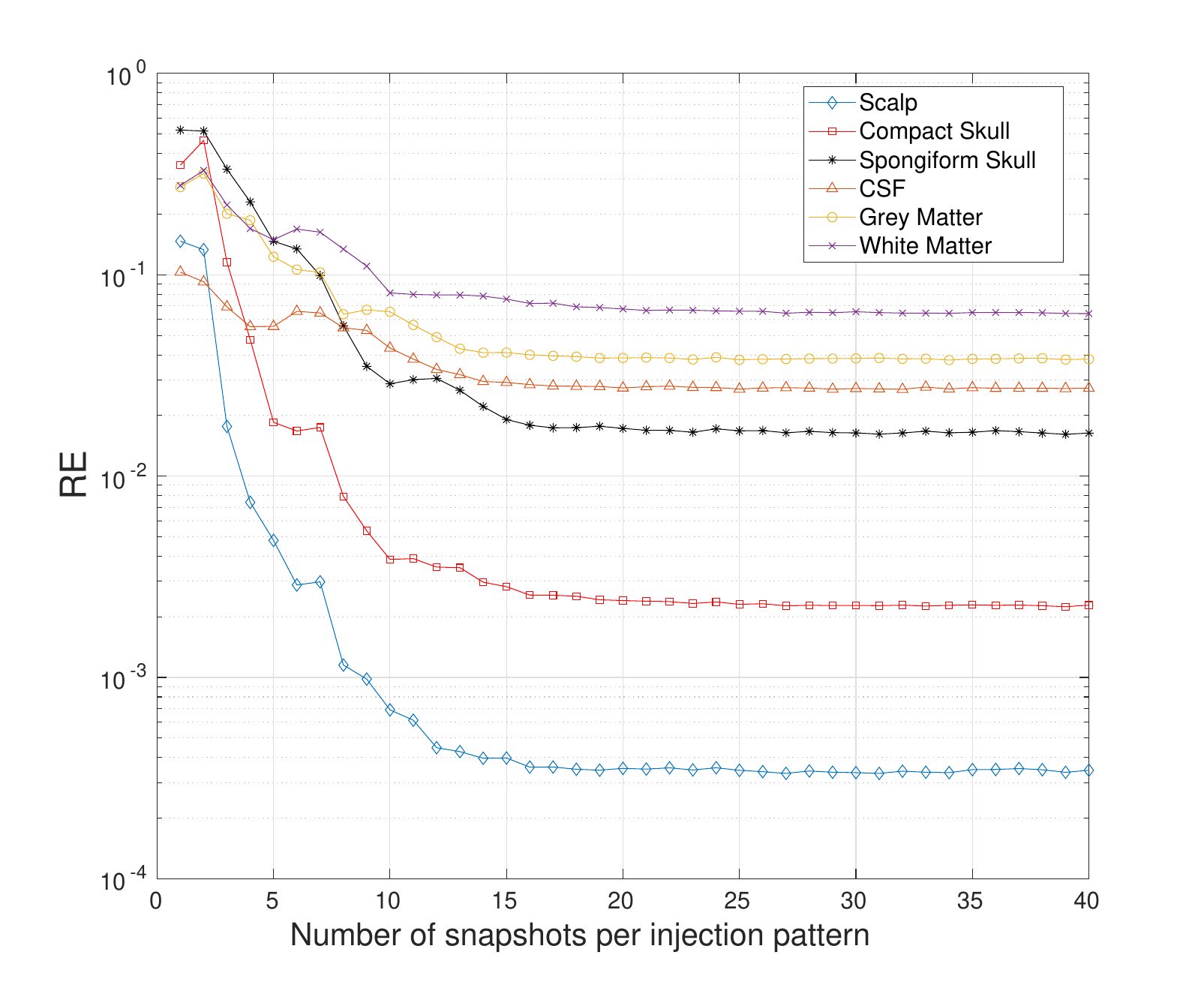}
    \caption{RE for each individual compartment as a function of snapshots across 100 samples and 132 injection pairs using the ROM-pEIT method.}
    \label{fig:inv_snap_ROM}
\end{figure}

\begin{figure}
    \centering
    \includegraphics[width=1\linewidth, trim=25pt 0pt 30pt 20pt, clip]{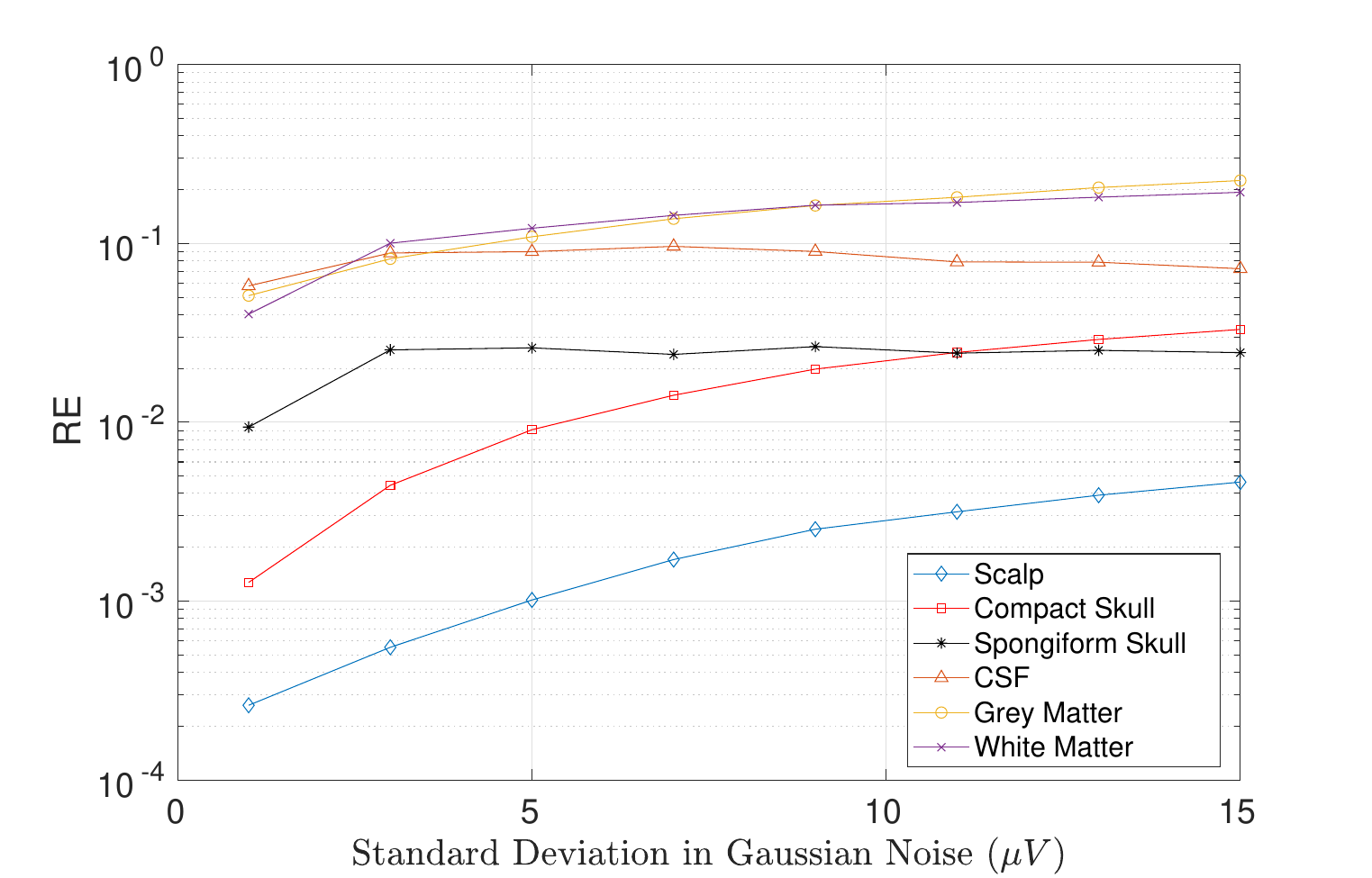}
    \caption{\response{RE in the conductivity estimations for each tissue as a function of the standard deviation of the Gaussian noise added to the measurements used in the IP.}}
    \label{fig:noise}
\end{figure}

\begin{figure}[t!]
    \centering
    \includegraphics[width=\linewidth,trim=35pt 20pt 60pt 40pt, clip]{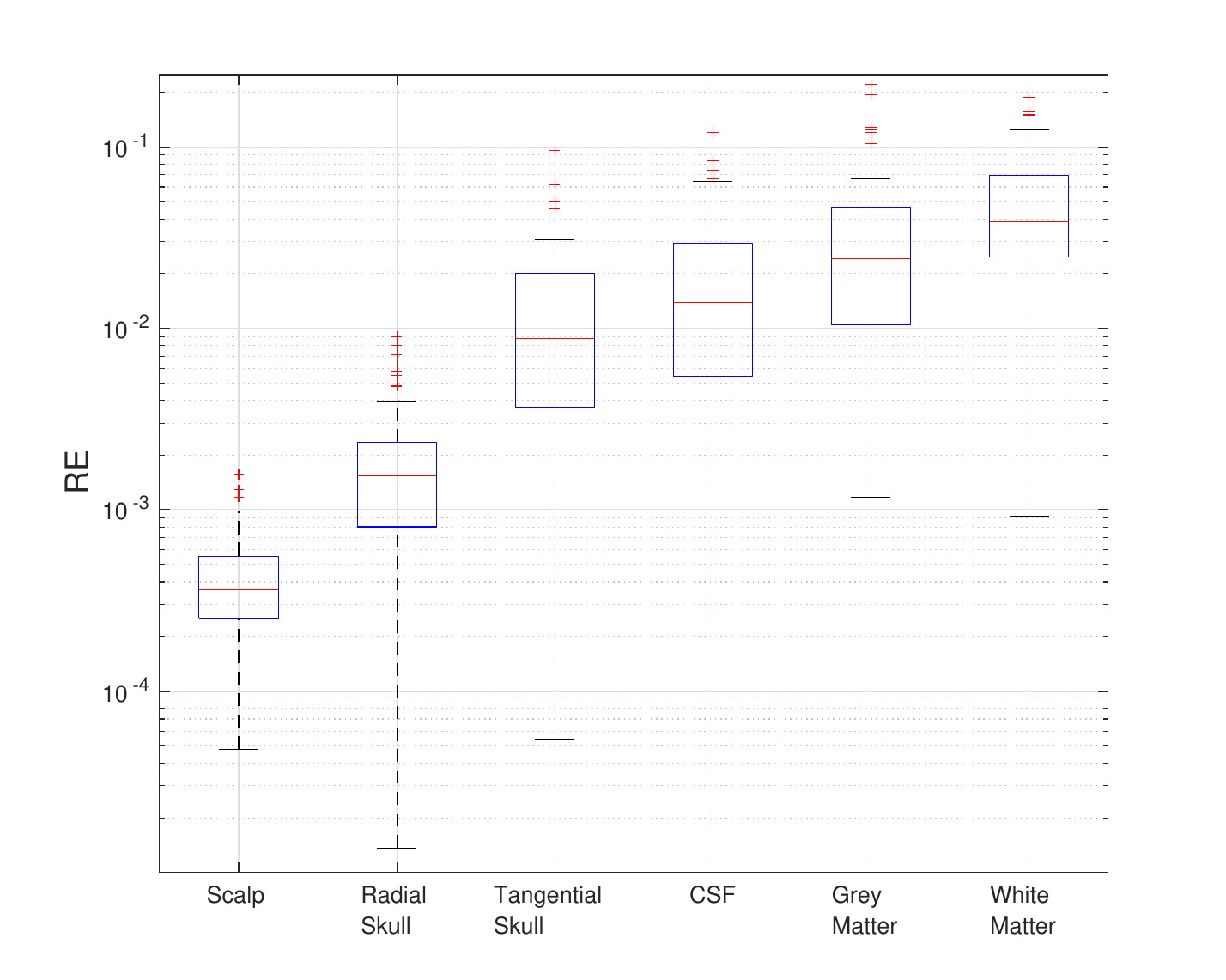}
    \caption{Sensitivity Analysis across 100 samples for the reduced basis anisotropic model. The estimations are for the full 6-parameter space using 30 snapshots for each electrode pair. Each box plot shows the estimation error in a single tissue that is labelled.}
    \label{fig:sense_anis}
\end{figure}

Displayed in Fig.~\ref{fig:inv_snap_all} is the average and maximum RE in the conductivity estimations for ROM and the traditional method across 10 samples and all electrode pairs. It can be seen that there are improvements in computational cost of the ROM-pEIT framework compared to the traditional method. This is shown for the first three compartments of the head model (scalp, compact skull and spongiform bone) and the scalp and spongiform bone separately. Focusing on the three compartment graph (Fig.~\ref{fig:inv_snap_all}c), we can see that the RE in the IP estimation averaged across compartments, injection patterns, samples of parameter space improves by nearly an order of magnitude, with the number of linear systems to solve reducing by an order of magnitude too. The maximum error for any injection pair for any sample is displayed in red crosses and also demonstrates an improvement over the average of the traditional method.

The number of injection pairs removed from the traditional estimations due to erroneous results was approximately 30 for two of the samples and none for the rest. All injection pairs were preserved for the ROM-pEIT IPs.

It is useful to separate all of the conductivities to see which are contributing the most to the REs seen in Fig.~\ref{fig:inv_snap_all}c. The RE for the scalp is shown in Fig.~\ref{fig:inv_snap_all}a, where the improvement in computational effort due to the ROM-pEIT framework is most apparent with a reduction in systems to solve from 250 to 10 maintaining an order of magnitude improvement in RE. In Fig.~\ref{fig:inv_snap_all}b, we see that the traditional method cannot obtain a reliable estimate for the spongiform bone with the optimisation implementation used. However, the ROM-pEIT framework is able to estimate the conductivity of the spongiform bone down to an average RE of almost 1$\%$ and a maximum RE of 5$\%$.

As previously mentioned, the benefits of using ROM become most clear during a 6-parameter search where the IP can optimise for all compartments in the model. Fig.~\ref{fig:inv_snap_ROM} shows the average RE for ROM for all tissue compartments as a function of the number of snapshots used in the estimation.
The figure shows that with ROM and the optimisations it allows, the IP is able to estimate CSF, GM and WM in the brain to approximately a 3$\%$, 4$\%$, 7$\%$ RE, respectively. It is also worth noting that the coefficient of variation in the electrode estimations was between 0.001 and 0.1 for all tissues after 30 snapshots.

From Figs.~\ref{fig:inv_snap_all} and~\ref{fig:inv_snap_ROM} it is clear that the accuracy of the IP with ROM stops improving after 30 snapshots. Therefore, we chose to only train the anisotropic reduced model in Experiment 3 up to this number to perform the sensitivity analysis.

\response{Each traditional method function evaluation required approximately 250 s for each PCG followed by 1200 s for the full gradient calculation (consisting of 3 additional PCGs and 3 large matrix calculations). Each PCG in a greedy algorithm took a similar time as the traditional method plus 10 seconds for the overhead of calculating the bound and orthonormalising the solution to the transformation matrix. }

\response{An additional substantial speed-up was achieved in the greedy algorithm by utilising the reduced model at the previous iteration to provide an initial guess for the PCG method when solving for a snapshot. As snapshots are added to the reduced model, the initial guess improves, leading to faster PCG solutions. Practically, this means that the time taken for one snapshot is halved after approximately 40 snapshots.}

\response{Times varied substantially due to innate variability in compute nodes, even of the same species and differences in convergence speeds. However, we found that the traditional method took, on average, 30-40 hours to converge using 75 function evaluations while trivially parallelised on a cluster. Conversely, the greedy algorithm used to train the reduced model took only 1.5 hours to reach 40 snapshots. The resulting 132 ROM-pEIT IPs (one for each pair) took approximately 20 seconds to complete in series on a single compute node.}

\response{Fig.~\ref{fig:noise} shows the response of the IP estimations for each tissue to varying intensities of noise, while utilising the interior-point optimisation and ROM-pEIT FPs. As expected, the RE in most tissues increase as noise increases. Of particular interest, is the RE in the spongiform bone, which remains stable under noise that is 1.5 orders of magnitude greater than the noise obtained from real measurements~\cite{fernandez-corazza_skull_2018}.}

\subsubsection{Experiment 3 - Anisotropy}

\begin{figure}[t!]
    \centering
    \includegraphics[width=1\linewidth, trim=0pt 230pt 420pt 0pt, clip]{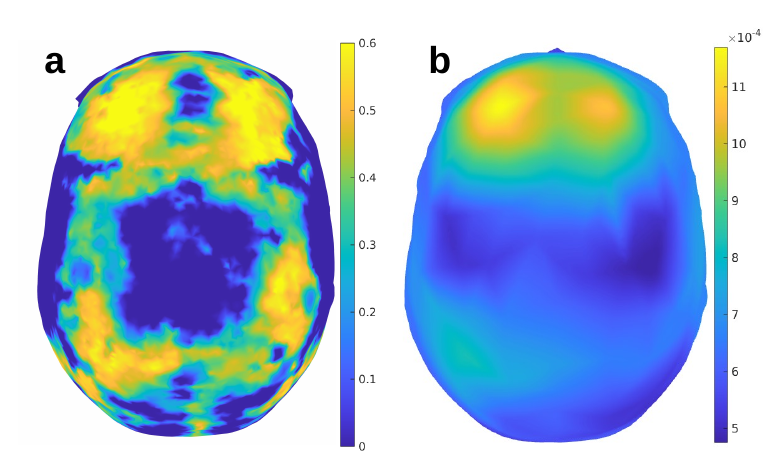}
    \caption{\response{\textbf{a)} Ratio of spongiform bone to compact bone that makes up a given point in the skull. \textbf{b)} Standard deviation (S/m) in the skull estimations in all electrode pairs plotted as a function of the reference electrode used and then interpolated over the skull.}}
    \label{fig:reference}
\end{figure}


\begin{table}[t]
\centering
\begin{tabular}{||c|c@{\hspace{1em}} c@{\hspace{1em}} c@{\hspace{1em}} c@{\hspace{1em}} c||}
 \cline{2-6}
  \multicolumn{1}{c|}{} & Scalp & Skull & CSF & GM & WM\\ [0.5ex]
  \hline
 True & 369 & (5.9/38) & 1626 & 421 & 134 \\ 
 Est. & 362$\ \pm\ $22 & 8.7$\ \pm\ $1.2 & 1602$\ \pm\ $157 & 325$\ \pm\ $98 & 101$\ \pm\ $25 \\ 
 \hline
\end{tabular}
\caption{\response{Ground truth and estimated conductivities in the situation of a missing spongiform bone compartment. Units in mS/m.}}
\label{table:no_marrow}
\end{table}

The results of the sensitivity analysis described in Section~\ref{method:anis} are displayed in Fig.~\ref{fig:sense_anis}. From this analysis we can see that the framework presented is sensitive to the tangential and radial conductivity components of the skull while remaining sensitive to the inner compartments. 
We found that, for 30 snapshots, the mean $\Delta_{RE}(\boldsymbol \sigma)$ was approximately $10^{-2}$ and the mean RE$(\boldsymbol \sigma)$ was $10^{-4}$.



\subsubsection{Experiment 4 - Response to Reference Choice}

\response{Fig.~\ref{fig:reference}a displays the ratio of spongiform bone to compact bone in the skull from 0 to 1. This shows the distribution of the spongiform bone within the skull. Plotted across the skull in Fig.~\ref{fig:reference}b are the standard deviations in the estimations for the skull compartment across electrode pairs as a function of the reference electrode used. \respm{In other words, the colour at each point on the skull represents the variability in 106 IP skull estimations if the reference electrode were placed over that point.} Note that the bright yellow areas clearly correlate with the spongiform bone information that is missing, indicating that when the reference electrode is close to a modelling inaccuracy, the estimations become less stable. The generation of Fig.~\ref{fig:reference}b required over 14,000 IPs and at least 10 million FPs, needing only 30 minutes to complete on a single PC (Intel(R) Core(TM) i5-6500 CPU at 3.20GHz). Within reason, this analysis could not be possible without ROM-pEIT. \par Table~\ref{table:no_marrow} contains the ground truth conductivities for the synthetic measurements and the estimated conductivities with standard deviations for all pairs when considering the reference electrode corresponding to the bright yellow spot in Fig.~\ref{fig:reference}. Note that the standard deviation is high in all compartments, as well as the skull, demonstrating the impact of incorrect skull modelling on the conductivity estimations.}

\subsubsection{Experiment 5 - Validation with Real Data}

\begin{table}[b]
\centering
\begin{tabular}{||c|c@{\hspace{1em}} c@{\hspace{1em}} c||}
 \hline
  Subject & ~~~Scalp (mS/m)~~~ & ~~~Compact~~~ & ~~~Spongiform~~~ \\ 
  & & bone (mS/m) & bone (mS/m) \\ [0.5ex] 
 \hline
 AM & 243$\ \pm\ $45 & 5.5$\ \pm\ $2.0 & 34$\ \pm\ $14 \\ 
 CA & 287$\ \pm\ $62 & 4.8$\ \pm\ $1.3 & 48$\ \pm\ $30 \\
 AA & 360$\ \pm\ $88 & 4.8$\ \pm\ $1.9 & 11$\ \pm\ $17 \\
 \hline
 AM & 214$\ \pm\ $58 & 5.5$\ \pm\ $2.5 & 195$\ \pm\ $128 \\ 
 CA & 277$\ \pm\ $64 & 4.1$\ \pm\ $1.1 & 225$\ \pm\ $103 \\
 AA & 345$\ \pm\ $98 & 4.8$\ \pm\ $2.1 & 174$\ \pm\ $143 \\
 \hline
\end{tabular}
\caption{\response{Conductivity estimations by subject data and compartment considering all intracranial compartments fixed using reduced model trained in the ranges from Table~\ref{table:conds} (top three rows) and the expanded spongiform range (bottom three rows).}}
\label{table:real_ests_fixed}
\end{table}

\begin{table}[b]
\centering
\begin{tabular}{||c|c@{\hspace{1em}} c@{\hspace{1em}} c||}
 \hline
  Subject & ~~~Scalp (mS/m)~~~ & ~~~Compact~~~ & ~~~Spongiform~~~ \\ 
  & & bone (mS/m) & bone (mS/m) \\ [0.5ex] 
 \hline
 AM & 237$\ \pm\ $42 & 6.4$\ \pm\ $2.5 & 35$\ \pm\ $14 \\ 
 CA & 271$\ \pm\ $64 & 6.0$\ \pm\ $2.7 & 45$\ \pm\ $29 \\
 AA & 337$\ \pm\ $94 & 5.9$\ \pm\ $2.8 & 18$\ \pm\ $27 \\
 \hline \hline
 & CSF (mS/m) & GM (mS/m) & WM (mS/m) \\ [0.5ex]
 \hline
 AM & 1455$\ \pm\ $145 & 303$\ \pm\ $278 & 118$\ \pm\ $72 \\ 
 CA & 1448$\ \pm\ $138 & 231$\ \pm\ $219 & 114$\ \pm\ $63 \\
 AA & 1433$\ \pm\ $219 & 219$\ \pm\ $259 & ~99$\ \pm\ $60 \\
 \hline
\end{tabular}
\caption{\response{Conductivity estimations of all compartments by subject data using reduced model trained in the ranges from Table~\ref{table:conds}.}}
\label{table:real_ests_unfixed}
\end{table}

\response{Table~\ref{table:real_ests_fixed} shows the average and standard deviation in the conductivity estimations obtained for the three participants considering the full conductivity ranges in Table~\ref{table:conds}. Similarly, the values estimated considering the expanded spongiform bone range can also be found in Table~\ref{table:real_ests_fixed}. These values are in strong agreement with the results obtained by Fern\'andez-Corazza et al. (2018) who obtained (for AM, CA and AA, respectively) scalp values of 249, 291, and 362 mS/m and compact skull values of 4.16, 4.22 and 4.25 mS/m with similar standard deviations~\cite{fernandez-corazza_skull_2018}. Table~\ref{table:real_ests_unfixed} contains the estimations for all tissues, found simultaneously, considering the ranges in Table~\ref{table:conds}. Note that in the interest of lowering the standard deviations of the estimations, we arbitrarily expanded the ranges used to train the reduced models for all tissue compartments and found no improvement. Similarly, we increased the number of snapshots taken to build the model up to 130 and still found no improvement over estimations made with 40 snapshots. \par \respm{It took approximately two hours to train each of the models up to 40 snapshots and 10 seconds to run all IPs for each subject. This is much faster than the estimations obtained by Fern\'andez-Corazza et al. (2018), which took days.}}


\section{Discussion}
\label{discussion}

We have presented a framework for the solution of the pEIT-FP using ROM, where we have demonstrated a significant reduction in computational expense, resulting in a framework at least 30 times faster than that of the current state-of-the-art approach. \response{Similarly, we have shown that huge improvements can be achieved in conductivity estimations for all tissues, many previously unreachable by pEIT in a reasonable time frame due to computational effort}. 

\subsection{Synthetic Data}

We have validated this approach experimentally by testing both methods on a realistic 6-layered head model to emulate typical use cases.
\response{In Fig.~\ref{fig:inv_snap_all} we compare the speed up of using the interior-point optimisation technique with ROM-pEIT and the traditional Newton method with full order system inversions. Fig.~\ref{fig:inv_snap_all} shows this in a scenario where the inner tissue conductivities are assumed to be known. A more realistic scenario would be that the inner tissue conductivities are unknown.}
In this instance, we found that after 200 full order systems solved the error in scalp estimations was half an order of magnitude higher than that achieved assuming the inner conductivities known (Fig.~\ref{fig:inv_snap_all}). We also found that the spongiform bone could not be estimated reasonably for the traditional algorithm.

For models that have been built from only T1-weighted MRI images, where segmenting spongiform bone in the skull accurately is not feasible, it has been shown to reduce errors in the EEG FP and IP when the anisotropy of the skull conductivity is considered~\cite{wolters_influence_2006}. In this context, ROM-pEIT also extends to such a situation.
Fig.~\ref{fig:sense_anis} also shows us that this IP is more sensitive to the radial conductivity than the tangential conductivity, which is consistent with reported findings~\cite{fernandez-corazza_analysis_2013}. 


\response{The interior-point optimisation afforded by the use of ROM-pEIT FPs was chosen based on its flexibility to handle large and small scale problems and the accuracy it provided in the IP estimations. In Experiment 2, this algorithm performed well, however, it is useful to assess its stability in the presence of different intensities of noise. Fig.~\ref{fig:noise} displays the results for this analysis. For the purposes of pEIT, this optimisation technique appears robust, however, other methods could be explored. With the speed of ROM-pEIT the analysis of other optimisation algorithms would become a less onerous task.}




\response{Furthermore, we provided evidence that the choice of reference electrode clearly effects the amount of standard deviation in the estimates substantially, as shown in Fig~\ref{fig:reference}b. This value is most affected over areas of modelling inaccuracy and vary by as much as 1.5 times the average. Fig~\ref{fig:reference}b demonstrates that errors in the potential at the reference electrode in the FP are propagated to all other electrodes, resulting in larger standard deviations in the estimations of each IP. Therefore, regardless of re-referencing scheme, the original reference can have a substantial impact when modelling inaccuracies are present.}

\response{This observation has wide implications in the field of EEG, TES and pEIT given that the reference electrode is often fixed in commercial electrode arrays to the centroparietal midline~\cite{hari_meg-eeg_2017}. Therefore, we suggest this standard practice be revisited to allow more flexible control of the position of the reference electrode.}

\response{It is worth noting that ROM-pEIT is useful for a pEIT setup with any number of electrodes. Here, we have provided a sensitivity analysis of ROM-pEIT applied to a typical pEIT setup, however, to further assess the validity of this methodology under different pEIT conditions, this analysis should be extended. This extension would include the impact of electrode position error, different sensor layouts and numbers, and contact impedance variability. These parameters are beyond the scope of this paper, however, they will be subject to further investigation.}

\subsection{Real Data}


\response{The ROM-pEIT framework has demonstrated strong agreement with the traditional method for the real pEIT data from three subjects~\cite{fernandez-corazza_skull_2018}.  It was expected that, given the additional freedom of the other compartments being estimated, the standard deviation in the estimations across the electrode patterns may reduce thanks to the entire model becoming more individualised, therefore removing errors introduced by incorrectly assigned conductivities. In this context the ROM-pEIT framework has proven in simulated data that searching the parameter space of this dimensionality is trivial. However, as seen in Tables~\ref{table:real_ests_fixed}-~\ref{table:real_ests_unfixed}, the coefficient of variation of the spongiform bone, CSF, GM and WM all remained high. When the ranges were expanded, the coefficient of variation in all tissues remained approximately the same.}


\response{Combined with the analysis of the reference placement and the importance of anatomically correct head models, confirmed in Table~\ref{table:no_marrow}, this leads us to hypothesise that a standard modelling assumption could be causing the variability in the estimations of a single subject. Given the standard deviations in the estimations of all compartments, the assumption of a homogeneous scalp layer could be playing a role. This may be reasonable to challenge due to its complex structure~\cite{ellis_surgical_2014}, role in EEG source localisation~\cite{vorwerk_global_2024} and fat content~\cite{ramon_influence_2006}.}

\response{From these observations, an investigation into the common modelling assumption of a single homogeneous scalp layer is warranted, and could have wide implications for the field of EEG, which relies heavily on these models. Crucially, this \textit{strongly} emphasises the need of pEIT and specifically the ROM-pEIT framework to challenge modelling assumptions. Without these tools, investigations of this nature would be incredibly taxing. We believe the framework we present is an essential tool for researchers in this context.}

\subsection{Related Work}

Some efforts have been made to avoid the computational expense of EIT while retrieving subject-specific conductivity values. Akalin Acar et al. (2016)~\cite{akalin_acar_simultaneous_2016}  and Costa et al.~(2017)~\cite{costa_skull_2017} demonstrated techniques for the simultaneous estimation of the conductivity of the skull, modelled a single compartment, and the location of the source of electrical activity. Others have used a pre-calibration technique for combined EEG and MEG where an initial conductivity value for the skull is given and then tuned before the source localisation by using somatosensory evoked potential data~\cite{wolters_combined_2010,aydin_combining_2014,antonakakis_inter-subject_2020}. However, these techniques have only been demonstrated for estimating the skull and brain conductivities. Moreover, the method presented in~\cite{akalin_acar_simultaneous_2016}, which uses only EEG data, requires computational effort to converge, reported to be in the order of days by the authors. ROM-pEIT allows all compartments to be estimated simultaneously in a reasonable time frame.

The computational costs of ESI related methods become particularly prohibitive when performing sensitivity analyses, where effects of conductivity uncertainty in specific head tissues is explored. One way this problem has been circumvented is through the use of generalized Polynomial Chaos (gPC) expansions, where a result distribution is described by a linear combination of multivariate orthogonal polynomial basis functions and corresponding coefficients~\cite{schmidt_impact_2015,vorwerk_influence_2019}. Similarly to ROM, this method involves the calculation of the model output at multiple points on a sparse grid with specific parameters required to weight the coefficients. This technique was utilised by Schmidt et al. (2015)~\cite{schmidt_impact_2015} for a sensitivity analysis in TES and by Vorwerk et al. (2019)~\cite{vorwerk_influence_2019} in EEG. Generalised PC has also been used for a conductivity uncertainty analysis in transcranial magnetic stimulation (TMS) and TES by Saturnino et al. (2019)~\cite{saturnino_principled_2019}. Although resulting in an essential reduction in computational effort for these experiments, they still required the evaluations of the full FP at hundreds of points in parameter space for gPC convergence. The framework we present requires only a few dozen full order FP evaluations to reach a low RE in the FPs and IPs.

A closely related work by Maksymenko et al. (2020) also demonstrates a reduced order technique for fast solutions of the EEG FP~\cite{maksymenko_fast_2020}. Similarly, this framework used a set of full-order solutions at points in parameter space chosen via a greedy algorithm. This model could generate approximate lead field matrices for any conductivity set in parameter space very rapidly. There are, however, some notable differences between this framework and the one that we present in this work. Mainly, the implementation differs, where the former is applied to the EEG FP and solved using the Boundary Element Method with a small number of nodes in a model with 3 tissue compartments. Although it is suggested that it could equally be applied to FEM, this is not shown. \response{Given the implementation in EEG, the bound in Maksymenko et al. (2020) is also derived used different parameters and cannot be compared to that obtained with ROM.}

Similarly, work by Codecasa et al. (2016)~\cite{codecasa_fast_2016} has merged the techniques of ROM and gPC to perform an uncertainty analysis in TMS, where the model order reduction is used to guide the selection of the conductivity samples used for the polynomial chaos expansion. This work resulted in a significant speed up over gPC with regression, demonstrating the power of reduced order model techniques. There are a few differences in our work that make it distinguishable from this, such as a bound on the approximation error, application to pEIT, and the investigation of 3 additional tissues (scalp, compact skull and spongiform skull).

It is worth mentioning that for studies involving gPC, where a model is trained using hundreds of support points, all were sensitivity analyses. Due to the nature of this work it is essential to have a highly trained model. However, for personalised conductivity field reconstruction, there is more interest in reducing the time from measurement to result. This is one of the strengths of ROM-pEIT. Shown in Figs.~\ref{fig:inv_snap_all} and~\ref{fig:inv_snap_ROM}, only ${30-40}$ support points per injection pattern are required for accurate estimations in all tissues.


\response{McCann et al. (2019) showed that spongiform bone varies between subjects and measurement techniques and that few attempts have been made to measure the conductivitity of this tissue \textit{in vivo}. In addition to this and the work by Fern\'andez-Corazza et al. (2018), Aydin et al. (2014) used a pre-EEG calibration method to obtain a value of 8.4 mS/m~\cite{aydin_combining_2014}. Clearly, \textit{in vivo} measurements of this tissue remain challenging. Reasons for this have previously included the computational burden, however, with ROM-pEIT, we demonstrate that this could also be due to modelling imperfections. Furthermore, to the best of our knowledge, only two studies on non-invasive $in \ vivo$ estimations of the CSF are present in the literature, with large errors in the estimations reported~\cite{ouypornkochagorn_vivo_2019,ferree_regional_2000}.}

\response{It is worth emphasising the difference between imaging EIT and pEIT. For the imaging modality of EIT, dimensionality reduction techniques have been explored in the form of basis constraints~\cite{vauhkonen_electrical_1997} and autoencoders~\cite{seo_learning-based_2019}, amongst others. Fundamentally, these approaches are tuned towards imaging EIT, which is a different type of problem to parametric EIT, distinguished by how ill-posed the IP is. For imaging, the conductivity at each pixel/voxel is reconstructed from only a handful of electrode measurements, whereas, in parametric EIT, only a handful of conductivities are estimated from as many as 256 electrode measurements. In the case of both aforementioned techniques, an approximate solution manifold is made, similarly to ROM. However, with basis constraints, the support points are hand selected, whereas, with autoencoders, the number of support points required is over 20,000 for a 16-channel system. The ill-posed nature of imaging EIT also requires stabilisation techniques such as Tikhonov regularisation~\cite{vauhkonen_tikhonov_1998}. This type of stabilisation does not apply here.}

\subsection{Future Work}

A key feature of this work is the certified upper bound on the error in the FP. Although it guarantees a maximum error for each snapshot number, its usefulness as a stopping criteria is limited given the sharpness of the bound. A further challenge is that drawing a connection between the error in the FP and IP is not trivial. However, from Figs.~\ref{fig:inv_snap_all} and~\ref{fig:inv_snap_ROM} it is clear that optimal performance was achieved after 30 snapshots. Additionally, when the error between the full-order and the RB signal becomes much smaller than the noise, eq.~\eqref{MLE} becomes approximately the norm of the noise over the electrodes. Relative to the measurements, that becomes approximately $4-5\times10^{-6}$ RE($\boldsymbol \sigma$) (for the noise we have used), which on Fig.~\ref{fig:bound} corresponds to about 30 snapshots, as observed in the IP. For our head model, this connects the observations in the FPs and IPs and therefore we suggest 30 snapshots as the optimal number and this can serve as a stopping criteria. 
\response{However, we emphasise that this choice in snapshot number could change depending on the head model discritisation, level of noise in the measurements, and the number of conductivities to estimate (as this will affect the dimensionality of the parameter space). When applied to the head models of the real participants, we found no change in estimation stability past 40 snapshots. This further supports the idea that we should challenge our modelling assumptions. Nevertheless, a full analysis to characterise this value is due.}

A time penalty incurred by ROM is the computational cost associated with training the stability factor interpolant during the offline phase, requiring multiple solutions to a generalised eigenvalue problem.
This process takes approximately 15-30 minutes per problem (for an Intel Xeon CPU at 2.8~GHz for our model) and can be parallelised on a cluster. The interpolant generated is source vector independent, and therefore can be used for all electrode pairs. Although small in comparison to the training for ROM and the traditional method, this should still be considered as part of the offline training process. There exists some techniques that minimise the computational load of this stage such as greedy algorithms to reduce the number of interpolation points needed~\cite{manzoni_heuristic_2015}. Exploring these optimisations of the framework will be work for the future. Further, we've found that interpolating between these points in a 6-dimensional space is a non-trivial task due to the complexity of the resulting manifold and the possible noise in the interpolation data. We found that the use of too many randomly selected interpolation points led to over-fitting and consequently a poor interpolation. The more conservative strategies suggested by Manzoni et al. (2015)~\cite{manzoni_heuristic_2015} may help tackle this class of problem and this shall be explored in future work.

In this framework, we use the L2-norm in both the $\Delta_{RE}(\boldsymbol \sigma)$  and the projection due to its ease of implementation. However, an equally valid $\Delta_{RE}(\boldsymbol \sigma)$  can also be calculated using the norm of the space containing the solution~\cite{quarteroni_reduced_2016}. The solution to the variational formulation of the problem can be found in an appropriate quotient Hilbert space, equipped with a norm that can be used for this task~\cite{somersalo_existence_1992}.
Modifying our framework to utilise this norm may improve the sharpness of the bound.

McCann et al. (2022) also investigated the effect of sutures on the EEG FP and IP and found that omission of the sutures from a head model led to significant source localisation errors~\cite{mccann_impact_2022}. It is unclear how the inclusion of sutures in a realistic head model may affect the training of the reduced order model, however this should be considered in future models. Moreover, with the possibility of estimating inner tissue compartments, the impact of including sutures on the estimation of the inner compartments could be assessed.

TES has been shown to produce a greater intensity and focality of the electric current at a point of interest when highly accurate head models are considered~\cite{wagner_investigation_2014} and optimal injection patterns are generated~\cite{sadleir_target_2012}. ROM could reduce the time constraints involved and in an online process estimate the conductivities and optimal injection patterns together almost instantly. Future work could involve producing a pipeline for TES such that the number of measurements taken from the patient are kept to a minimum.


One artefact of the training noticed was the loss of orthogonality in the transformation matrix after approximately 150 snapshots. This could be due to numerical errors introduced into the Gram-Schmidt orthonormalisation. 
We use the classical Gram-Schmidt process in this work, however, a well known and more numerically stable method called the modified Gram-Schmidt method could also be used~\cite{bjorck_numerics_1994}. Other numerically stable implementations of the Gram-Schmidt process have been developed and these may be explored in the future~\cite{carson_block_2022}.


In conclusion, this new framework embodies a fresh approach to pEIT that will change its accessibility and reliability, recasting its role in the generation of personalised realistic head models used for ESI methods.


The software developed for this research can be found here: 

\noindent https://github.com/09nwalkerm/ROMpEIT. 

\section*{Acknowledgements}

We would like to thank Carlos H. Muravchik for helpful discussions and comments. This work is supported by the UKRI AIMLAC CDT, funded by grant EP/S023992/1. For the purpose of open access, the author(s) has applied a Creative Commons Attribution (CC BY) licence to any Author Accepted Manuscript version arising.

\section*{References}
 
\bibliographystyle{ieeetr}

\begin{thebibliography}{10}

\bibitem{michel_eeg_2019}
C.~M. Michel and D.~Brunet, ``{EEG} {Source} {Imaging}: {A} {Practical} {Review} of the {Analysis} {Steps},'' {\em Frontiers in Neurology}, vol.~10, p.~325, 2019.

\bibitem{mccann_impact_2022}
H.~McCann and L.~Beltrachini, ``Impact of skull sutures, spongiform bone distribution, and aging skull conductivities on the {EEG} forward and inverse problems,'' {\em Journal of Neural Engineering}, vol.~19, p.~016014, 2022.

\bibitem{vatta_realistic_2010}
F.~Vatta et al., ``Realistic and {Spherical} {Head} {Modeling} for {EEG} {Forward} {Problem} {Solution}: {A} {Comparative} {Cortex}-{Based} {Analysis},'' {\em Computational Intelligence and Neuroscience}, vol.~2010, pp.~1--11, 2010.

\bibitem{hunold_review_2023}
A.~Hunold et al., ``Review of individualized current flow modeling studies for transcranial electrical stimulation,'' {\em Journal of Neuroscience Research}, vol.~101, pp.~405--423, 2023.

\bibitem{puonti_accurate_2020}
O.~Puonti et al., ``Accurate and robust whole-head segmentation from magnetic resonance images for individualized head modeling,'' {\em NeuroImage}, vol.~219, p.~117044, 2020.

\bibitem{mccann_variation_2019}
H.~McCann, G.~Pisano, and L.~Beltrachini, ``Variation in {Reported} {Human} {Head} {Tissue} {Electrical} {Conductivity} {Values},'' {\em Brain Topography}, vol.~32, pp.~825--858, 2019.

\bibitem{haueisen_influence_2002}
J.~Haueisen et al., ``The {Influence} of {Brain} {Tissue} {Anisotropy} on {Human} {EEG} and {MEG},'' {\em NeuroImage}, vol.~15, pp.~159--166, 2002.

\bibitem{pohlmeier_influence_1997}
R.~Pohlmeier et al., ``The influence of skull-conductivity misspecification on inverse source localization in realistically shaped finite element head models,'' {\em Brain Topography}, vol.~9, pp.~157--162, 1997.

\bibitem{vorwerk_influence_2019}
J.~Vorwerk et al., ``Influence of {Head} {Tissue} {Conductivity} {Uncertainties} on {EEG} {Dipole} {Reconstruction},'' {\em Frontiers in Neuroscience}, vol.~13, p.~531, 2019.

\bibitem{wolters_influence_2006}
C.~Wolters et al., ``Influence of tissue conductivity anisotropy on {EEG}/{MEG} field and return current computation in a realistic head model: {A} simulation and visualization study using high-resolution finite element modeling,'' {\em NeuroImage}, vol.~30, pp.~813--826, 2006.

\bibitem{wagner_investigation_2014}
S.~Wagner et al., ``Investigation of {tDCS} volume conduction effects in a highly realistic head model,'' {\em Journal of Neural Engineering}, vol.~11, p.~016002, 2014.

\bibitem{mccann_does_2021}
H.~McCann and L.~Beltrachini, ``Does participant’s age impact on {tDCS} induced fields? {Insights} from computational simulations,'' {\em Biomedical Physics \& Engineering Express}, vol.~7, p.~045018, 2021.

\bibitem{holder_electrical_2004}
D.~Holder, ed., {\em Electrical {Impedance} {Tomography}}.
\newblock CRC Press, 0~ed., 2004.

\bibitem{simini_effects_2016}
M.~Fernández-Corazza et al., ``Effects of {Head} {Model} {Inaccuracies} on {Regional} {Scalp} and {Skull} {Conductivity} {Estimation} {Using} {Real} {EIT} {Measurements},'' in {\em {II} {Latin} {American} {Conference} on {Bioimpedance}} (F.~Simini and P.~Bertemes-Filho, eds.), vol.~54, pp.~5--8, 2016.
\newblock Series Title: IFMBE Proceedings.

\bibitem{fernandez-corazza_skull_2018}
M.~Fernández-Corazza et al. ``Skull {Modeling} {Effects} in {Conductivity} {Estimates} {Using} {Parametric} {Electrical} {Impedance} {Tomography},'' {\em IEEE Transactions on Biomedical Engineering}, vol.~65, pp.~1785--1797, 2018.

\bibitem{fernandez-corazza_analysis_2013}
M.~Fernández-Corazza et al., ``Analysis of parametric estimation of head tissue conductivities using {Electrical} {Impedance} {Tomography},'' {\em Biomedical Signal Processing and Control}, vol.~8, pp.~830--837, 2013.

\bibitem{quarteroni_reduced_2016}
A.~Quarteroni, A.~Manzoni, and F.~Negri, {\em Reduced {Basis} {Methods} for {Partial} {Differential} {Equations}}, vol.~92 of {\em {UNITEXT}}.
\newblock Cham: Springer International Publishing, 2016.

\bibitem{somersalo_existence_1992}
E.~Somersalo, M.~Cheney, and D.~Isaacson, ``Existence and {Uniqueness} for {Electrode} {Models} for {Electric} {Current} {Computed} {Tomography},'' {\em SIAM Journal on Applied Mathematics}, vol.~52, no.~4, pp.~1023--1040, 1992.

\bibitem{fernandez-corazza_estimation_2011}
M.~Fernández-Corazza, N.~von Ellenrieder, and C.~H. Muravchik, ``Estimation of electrical conductivity of a layered spherical head model using electrical impedance tomography,'' {\em Journal of Physics: Conference Series}, vol.~332, p.~012022, 2011.

\bibitem{vauhkonen_three-dimensional_1999}
P.~Vauhkonen et al., ``Three-dimensional electrical impedance tomography based on the complete electrode model,'' {\em IEEE Transactions on Biomedical Engineering}, vol.~46, pp.~1150--1160, 1999.

\bibitem{kuo-sheng_cheng_electrode_1989}
{Kuo-Sheng Cheng} et al., ``Electrode models for electric current computed tomography,'' {\em IEEE Transactions on Biomedical Engineering}, vol.~36, pp.~918--924, 1989.

\bibitem{vauhkonen_fixed-lag_2001}
P.~J. Vauhkonen, M.~Vauhkonen, and J.~P. Kaipio, ``Fixed-lag smoothing and state estimation in dynamic electrical impedance tomography,'' {\em International Journal for Numerical Methods in Engineering}, vol.~50, pp.~2195--2209, 2001.

\bibitem{aubert-broche_new_2006}
B.~Aubert-Broche, A.~C. Evans, and L.~Collins, ``A new improved version of the realistic digital brain phantom,'' {\em NeuroImage}, vol.~32, pp.~138--145, 2006.

\bibitem{elman_finite_2014}
H.~Elman, D.~Silvester, and A.~Wathen, {\em Finite {Elements} and {Fast} {Iterative} {Solvers}: with {Applications} in {Incompressible} {Fluid} {Dynamics}}.
\newblock Oxford University PressOxford, 2~ed., 2014.

\bibitem{beltrachini_analytical_2019}
L.~Beltrachini, ``The analytical subtraction approach for solving the forward problem in {EEG},'' {\em Journal of Neural Engineering}, vol.~16, p.~056029, 2019.

\bibitem{pursiainen_complete_2012}
S.~Pursiainen, F.~Lucka, and C.~H. Wolters, ``Complete electrode model in {EEG}: relationship and differences to the point electrode model,'' {\em Physics in Medicine and Biology}, vol.~57, pp.~999--1017, 2012.

\bibitem{dennis_numerical_1996}
J.~E. Dennis and R.~B. Schnabel, {\em Numerical {Methods} for {Unconstrained} {Optimization} and {Nonlinear} {Equations}}.
\newblock Society for Industrial and Applied Mathematics, 1996.

\bibitem{dannhauer_modeling_2011}
M.~Dannhauer et al., ``Modeling of the human skull in {EEG} source analysis,'' {\em Human Brain Mapping}, vol.~32, pp.~1383--1399, 2011.

\bibitem{hari_meg-eeg_2017}
R.~Hari and A.~Puce, {\em {MEG}-{EEG} {Primer}}.
\newblock Oxford University Press, 1~ed., Mar. 2017.

\bibitem{chella_impact_2016}
F.~Chella et al., ``Impact of the reference choice on scalp {EEG} connectivity estimation,'' {\em Journal of Neural Engineering}, vol.~13, p.~036016, 2016.

\bibitem{liu_estimating_2015}
Q.~Liu et al., ``Estimating a neutral reference for electroencephalographic recordings: the importance of using a high-density montage and a realistic head model,'' {\em Journal of Neural Engineering}, vol.~12, p.~056012, 2015.

\bibitem{yao_method_2001}
D.~Yao, ``A method to standardize a reference of scalp {EEG} recordings to a point at infinity,'' {\em Physiological Measurement}, vol.~22, pp.~693--711, 2001.

\bibitem{qianqian_fang_tetrahedral_2009}
{Qianqian Fang} and D.~A. Boas, ``Tetrahedral mesh generation from volumetric binary and grayscale images,'' in {\em 2009 {IEEE} {International} {Symposium} on {Biomedical} {Imaging}: {From} {Nano} to {Macro}}, (Boston, MA, USA), pp.~1142--1145, IEEE, 2009.

\bibitem{ellis_surgical_2014}
H.~Ellis and V.~Mahadevan, ``The surgical anatomy of the scalp,'' {\em Surgery (Oxford)}, vol.~32, pp.~e1--e5, 2014.

\bibitem{vorwerk_global_2024}
J.~Vorwerk, C.~H. Wolters, and D.~Baumgarten, ``Global sensitivity of {EEG} source analysis to tissue conductivity uncertainties,'' {\em Frontiers in Human Neuroscience}, vol.~18, p.~1335212, 2024.

\bibitem{ramon_influence_2006}
C.~Ramon, P.~H. Schimpf, and J.~Haueisen, ``Influence of head models on {EEG} simulations and inverse source localizations,'' {\em BioMedical Engineering OnLine}, vol.~5, p.~10, 2006.

\bibitem{akalin_acar_simultaneous_2016}
Z.~Akalin~Acar, C.~E. Acar, and S.~Makeig, ``Simultaneous head tissue conductivity and {EEG} source location estimation,'' {\em NeuroImage}, vol.~124, pp.~168--180, 2016.

\bibitem{costa_skull_2017}
F.~Costa et al., ``Skull {Conductivity} {Estimation} for {EEG} {Source} {Localization},'' {\em IEEE Signal Processing Letters}, vol.~24, pp.~422--426, 2017.

\bibitem{wolters_combined_2010}
C.~H. Wolters et al., ``Combined {EEG}/{MEG} source analysis using calibrated finite element head models,'' {\em Biomedizinische Technik/Biomedical Engineering. Rostock, Germany: Walter de Gruyter}, vol.~55, no.~Suppl 1, pp.~64--68, 2010.
\newblock Publisher: Citeseer.

\bibitem{aydin_combining_2014}
{\" U}.~Aydin et al., ``Combining {EEG} and {MEG} for the {Reconstruction} of {Epileptic} {Activity} {Using} a {Calibrated} {Realistic} {Volume} {Conductor} {Model},'' {\em PLoS ONE}, vol.~9, p.~e93154, 2014.

\bibitem{antonakakis_inter-subject_2020}
M.~Antonakakis et al., ``Inter-{Subject} {Variability} of {Skull} {Conductivity} and {Thickness} in {Calibrated} {Realistic} {Head} {Models},'' {\em NeuroImage}, vol.~223, p.~117353, 2020.

\bibitem{schmidt_impact_2015}
C.~Schmidt et al., ``Impact of uncertain head tissue conductivity in the optimization of transcranial direct current stimulation for an auditory target,'' {\em Journal of Neural Engineering}, vol.~12, p.~046028, 2015.

\bibitem{saturnino_principled_2019}
G.~B. Saturnino et al., ``A principled approach to conductivity uncertainty analysis in electric field calculations,'' {\em NeuroImage}, vol.~188, pp.~821--834, 2019.

\bibitem{maksymenko_fast_2020}
K.~Maksymenko, M.~Clerc, and T.~Papadopoulo, ``Fast {Approximation} of {EEG} {Forward} {Problem} and {Application} to {Tissue} {Conductivity} {Estimation},'' {\em IEEE Transactions on Medical Imaging}, vol.~39, pp.~888--897, 2020.

\bibitem{codecasa_fast_2016}
L.~Codecasa et al., ``Fast {MOR}-{Based} {Approach} to {Uncertainty} {Quantification} in {Transcranial} {Magnetic} {Stimulation},'' {\em IEEE Transactions on Magnetics}, vol.~52, pp.~1--4, 2016.

\bibitem{ouypornkochagorn_vivo_2019}
T.~Ouypornkochagorn and S.~Ouypornkochagorn, ``In {Vivo} {Estimation} of {Head} {Tissue} {Conductivities} {Using} {Bound} {Constrained} {Optimization},'' {\em Annals of Biomedical Engineering}, vol.~47, pp.~1575--1583, 2019.

\bibitem{ferree_regional_2000}
T.~Ferree, K.~Eriksen, and D.~Tucker, ``Regional head tissue conductivity estimation for improved {EEG} analysis,'' {\em IEEE Transactions on Biomedical Engineering}, vol.~47, pp.~1584--1592, 2000.


\bibitem{vauhkonen_electrical_1997}
M.~Vauhkonen et al., ``Electrical impedance tomography with basis constraints,'' {\em Inverse Problems}, vol.~13, pp.~523--530, 1997.

\bibitem{seo_learning-based_2019}
J.~K. Seo et al., ``A {Learning}-{Based} {Method} for {Solving} {Ill}-{Posed} {Nonlinear} {Inverse} {Problems}: {A} {Simulation} {Study} of {Lung} {EIT},'' {\em SIAM Journal on Imaging Sciences}, vol.~12, pp.~1275--1295, 2019.

\bibitem{vauhkonen_tikhonov_1998}
M.~Vauhkonenet al., ``Tikhonov regularization and prior information in electrical impedance tomography,'' {\em IEEE Transactions on Medical Imaging}, vol.~17, pp.~285--293, 1998.

\bibitem{manzoni_heuristic_2015}
A.~Manzoni and F.~Negri, ``Heuristic strategies for the approximation of stability factors in quadratically nonlinear parametrized {PDEs},'' {\em Advances in Computational Mathematics}, vol.~41, pp.~1255--1288, 2015.

\bibitem{sadleir_target_2012}
R.~J. Sadleir et al., ``Target {Optimization} in {Transcranial} {Direct} {Current} {Stimulation},'' {\em Frontiers in Psychiatry}, vol.~3, 2012.

\bibitem{bjorck_numerics_1994}
{\r A}.~Björck, ``Numerics of {Gram}-{Schmidt} orthogonalization,'' {\em Linear Algebra and its Applications}, vol.~197-198, pp.~297--316, 1994.

\bibitem{carson_block_2022}
E.~Carson et al., ``Block {Gram}-{Schmidt} algorithms and their stability properties,'' {\em Linear Algebra and its Applications}, vol.~638, pp.~150--195, 2022.


\end{thebibliography}

\end{document}